\documentclass[twocolumn]{IEEEtran}
\usepackage{graphics, subfigure, theorem, times, amsfonts, amsmath, amssymb, cite}
\usepackage{tikz, epic,eepic}
\usetikzlibrary{shapes,arrows}
\usepackage{pgfplots}
\usepackage{color}
\usepackage{hyperref,url}
\definecolor{darkblue}{rgb}{0,0.08,0.4} 
\definecolor{darkred}{rgb}{0.4,0.08,0} 
\input{mysymbol.sty}
\usepackage[margin=0.75in]{geometry}
\newtheorem{theorem}{Theorem}

\newtheorem{corollary}{Corollary}
\newtheorem{lemma}{Lemma}

{\itshape}{\rmfamily}
\newtheorem{assumption}{Assumption}

\theorembodyfont{\rm}
\newtheorem{remark}{\hspace{0pt}\bf Remark}

\def \forall {\text{for all\ }}
\def \TV {\text{TV}}

\usepackage{latexsym}
\usepackage{amscd, verbatim,mathtools}
\usepackage{hyperref,url,setspace}
\usepackage{algorithm,algorithmic}

\title{Distributed Fictitious Play for Optimal Behavior of Multi-Agent Systems with Incomplete Information}
\author{Ceyhun Eksin and Alejandro Ribeiro
\thanks{Work supported by the NSF award CAREER CCF-0952867 and the AFOSR MURI FA9550-10-1-0567. 
C. Eksin is with the School of Electrical and Computer Engineering and School of Biology at the Georgia Institute of Technology, Atlanta, GA 30332. 
A. Ribeiro is with the Department of Electrical and Systems Engineering, University of Pennsylvania, Philadelphia, PA 19104. Email: ceyhuneksin@gatech.edu.}}
\date{\today}
\begin{document}
\normalsize
\maketitle

\begin{abstract}
A multi-agent system operates in an uncertain environment about which agents have different and time varying beliefs that, as time progresses, converge to a common belief. A global utility function that depends on the realized state of the environment and actions of all the agents determines the system's optimal behavior. We define the asymptotically optimal action profile as an equilibrium of the potential game defined by considering the expected utility with respect to the asymptotic belief. At finite time, however, agents have not entirely congruous beliefs about the state of the environment and may select conflicting actions. This paper proposes a variation of the fictitious play algorithm which is proven to converge to equilibrium actions if the state beliefs converge to a common distribution at a rate that is at least linear. In conventional fictitious play, agents build beliefs on others' future behavior by computing histograms of past actions and best respond to their expected payoffs integrated with respect to these histograms. In the variations developed here histograms are built using knowledge of actions taken by nearby nodes and best responses are further integrated with respect to the local beliefs on the state of the environment. We exemplify the use of the algorithm in coordination and target covering games.
\end{abstract}

%
\section{Introduction}

Our model of a multi-agent autonomous system encompasses an underlying environment, knowledge about the state of the environment that the agents acquire, and a state dependent global objective that agents affect through their individual actions. The optimal action profile maximizes this global objective for the realized environment's state with the optimal action of an agent given by the corresponding action in the profile. The problem addressed in this paper is the determination of suitable actions when the probability distributions that agents have on the state of the environment are possibly different. These not entirely congruous beliefs result in mismatches between the action profiles that different agents deem to be optimal. As a consequence, when a given agent chooses an action to execute, it is important for it to reason about what the beliefs of other agents may be and what are the consequent actions that other agents may take. We propose a solution based on the construction of empirical histograms of past actions and the use of best responses to the utility expectation with respect to these histograms and the state belief. This algorithmic behavior is shown to be asymptotically optimal in the sense that if agents move towards a common belief, the actions they select are optimal with respect to the corresponding expected utility.

While the determination of optimal behavior in multi-agent systems can be considered from different perspectives, the categorization between systems with complete and incomplete information is most germane to this paper \cite{Shoham_Leyton_2008}. In systems of complete information the environment is either perfectly known or all agents have identical beliefs. In either case, agents can compute the optimal action profile, and determine and execute their corresponding optimal action. This local computation of global solutions is neither scalable nor robust but it can be used as an abstract definition of optimal behavior. This abstraction renders the problem equivalent to the development of distributed methodologies to solve optimization problems \cite{EksinRibeiro_2012,NedicOzdaglar,Tsitsiklisetal,chenSayed}, or, more generically, to the determination of Nash equilibria of multiplayer games \cite{Monderer_Shapley_1996a,Swenson_et_al_2014,Marden_et_al_2009,Shamma_Arslan_2005,Hart_2005,Young_2004}. When information is incomplete, the fact that agents have different beliefs implies that they may end up choosing competing actions even if they are intent on cooperating. In a sense, agents are competing against uncertainty, but the manifestation of that competition is in the form of conflicting interests arising from cooperating agents. In this inherent competition Bayesian Nash equilibria are the intrinsic mathematical formulation of optimal behavior \cite{Harsanyi_1968,EksinEtal13_b}. However, determination of these equilibria is computationally intractable except for games with simple beliefs and utilities \cite{Eksin_et_al_2013,Dekel_et_al_2004}. 

If determination of Bayesian equilibria is intractable, the development of approximate methods is necessary. In fact, determining game equilibria is also challenging in games of complete information. This has motivated the development of iterative methods to learn regular -- as opposed to Bayesian -- equilibrium actions \cite{Fudenberg_Levine_1998,Young_2004}. Of relevance to this paper is the fictitious play algorithm in which agents build beliefs on others' future behavior by computing histograms of past actions and best respond to their expected payoffs integrated with respect to these histograms \cite{Brown_1951}. When information is complete, fictitious play converges to equilibria of zero sum \cite{Fudenberg_Levine_1998}, some other specific games with two players \cite{Fudenberg_Kreps_1993}, and multiplayer games with aligned interests as determined by a potential function \cite{Monderer_Shapley_1996a}. Recent variations of fictitious play have been developed to expand the class of games for which equilibria can be computed \cite{Shamma_Arslan_2005,Fudenberg_Takahashi} and to guarantee convergence to equilibria with specific properties \cite{Arslan_et_al_2007, Marden_et_al_2009, Marden_Shamma_2012}. Recently, a variant of the fictitious play that operates in a distributed setting where agents observe relevant information from agents that are adjacent in a network is shown to converge in potential games \cite{Swenson_et_al_2014}. 

In this paper, we consider a network of agents with aligned interests. Agents have different and time varying beliefs on the state of the environment that, as time progresses, converge to a common belief. The asymptotic optimal behavior is therefore formulated as the Nash equilibria of the potential game defined by the expected utility with respect to the asymptotic belief. The goal is to design a distributed mechanism that converges to a Nash equilibrium of this asymptotic game (Section \ref{model_section}). The solutions that we propose are variations of the fictitious play algorithm that take into account the distributed nature of the multi-agent system and the fact that the state of the environment is not perfectly known. In a game of incomplete information, expected payoff computation in fictitious play consists of integrating the payoff with respect to both the local belief on the state of the environment and the local beliefs on the behavior of other agents. In a networked setting only local past histories can be available and agents need to reason about the behavior of non-neighboring agents based on past observations of its neighbors only. 

In potential games with symmetric payoffs, which are known to admit consensus Nash equilibria, we let agents share their actions with their neighbors, construct empirical histograms of the actions taken by neighbors, and best respond assuming that all agents follow the average population empirical distribution (Section \ref{distributed_fictitious_play_section}). This mechanism is shown to converge to a consensus Nash equilibrium when the convergence of individual beliefs on the state of the environment is at least of linear order (Theorem \ref{action_sharing_convergence}). When the potential game is not necessarily symmetric, agents share their empirical beliefs on the behavior of other agents with neighbors in addition to their own actions. Agents keep histograms on the behavior of others by averaging the histograms of neighbors with their own (Section \ref{histogram_sharing_section}). Convergence to a Nash equilibrium follows under the same linear convergence assumption for the beliefs on the state of the environment (Theorem \ref{histogram_sharing_convergence}). 

We numerically analyze the transient and asymptotic equilibrium properties of the algorithms in the beauty contest and the target covering games (Section \ref{numerical_examples}). In the beauty contest game, a team of robots tradeoffs between moving toward a target direction and moving in coordination with each other. In the target covering game, a team of robots coordinates to cover a given set of targets and receive payoffs from covering a target that is inversely proportional to the distance to their positions. We observe that Nash equilibrium strategies are successfully determined in both cases.

{\bf Notation:} For any finite set $X$, we use $\bigtriangleup(X)$ to denote the space of probability distributions over $X$. For a generic vector $x \in X^n$, $x_i$ denotes the $i$th element and $x_{-i}$ denotes the vector of elements of $x$ except the $i$th element, that is, $x_{-i} = (x_1,\dots,x_{i-1},x_{i+1},\dots,x_n)$. We use $\| \cdot \|$ to denote the Euclidean norm of a space.  $\bbone_n$ denotes a $n \times 1$ column vector of all ones.

%
\section{Optimal behavior of multi-agent systems with incomplete information} \label{model_section}

We consider a group of $n$ agents $i \in \ccalN:= \{1,\dots, n\}$ that play a stage game over a discrete time index $t$ with simultaneous moves and incomplete information. The important features of this game are the actions $a_{it}$ that agents take at time $t$, an underlying unknown state of the world $\theta$, local utility functions $u_i(a,\theta)$ that determine the payoffs that different agents receive when the group plays the joint action $a := \{a_1,\ldots, a_n\}$ and the state of the world is $\theta$, and time varying local beliefs $\mu_{it}$ that assign probabilities to different realizations of the state of the world. We assume that the state of the world is chosen by nature from a space $\Theta$ and that actions are chosen from a common, finite, and time invariant set so that we have $ a_{it}\in\ccalA :=\{1,\dots, m\}$ for all times $t$ and agents $i$. Local payoffs are then formally defined as functions $u_i: \ccalA^n \times\Theta\to\reals$. To emphasize the global dependence of local payoffs we write payoff values as $u_i(a,\theta)=u_i(a_i,a_{-i},\theta)$ where, we recall, $a_{-i}:=\{a_j\}_{j\neq i}$ collects the actions of all agents except $i$. We assume that the utility values $u_i(a,\theta)$ are finite for all actions $a$ and state realizations $\theta$. The beliefs $\mu_{it}\in\bigtriangleup(\Theta)$ are probability distributions on the space $\Theta$.

In general, we allow agent $i$ to maintain a mixed strategy $\sigma_i\in \bigtriangleup(\ccalA)$ defined as a probability distribution on the action space $\ccalA$ such that $\sigma_i(a_i)$ is the probability that $i$ plays action $a_i\in \ccalA$. The joint mixed strategy profile $\sigma:= \{\sigma_1,\dots, \sigma_n\}\in\bigtriangleup^n(\ccalA)$ is defined as the product distribution of all individual strategies and the mixed strategy of all agents except $i$ is written as $\sigma_{-i}:= \{\sigma_j\}_{j\neq i}\in\bigtriangleup^{n-1}(\ccalA)$. The utility associated with the joint mixed strategy profile $\sigma$ is then given by
\begin{equation} \label{eqn_expected_utility}
   u_i(\sigma, \theta) = u_i(\sigma_i, \sigma_{-i}, \theta)
                      = \sum_{ a \in \ccalA^n}u_i(a, \theta) \sigma(a).
\end{equation}
We further assume that there exists a {\it global} potential function $u: \ccalA^n \times \theta \mapsto \reals$ taking values $u(a,\theta)$ such that for all pairs of action profiles $a=\{a_i,a_{-i}\}$ and $a'=\{a_i',a_{-i}\}$, state realizations $\theta\in\Theta$, and agents $i$, the {\it local} payoffs satisfy
\begin{equation} \label{potential_game_definition}
    u_i(a_i, a_{-i},\theta) - u_i(a_i', a_{-i},\theta) 
        = u(a_i, a_{-i},\theta) - u(a_i', a_{-i},\theta) .
\end{equation}
The existence of the potential function $u$ is a statement of aligned interests because for a given $\theta$ the joint action that maximizes $u$ is a pure Nash equilibrium strategy of the game defined by the $u_i$ utilities \cite{Monderer_Shapley_1996a}. The motivation for considering a potential game is to model a system in which agents play to achieve the game equilibrium in a distributed manner and end up finding the action $a$ that would be selected by a central coordination agent that maximizes the global payoff $u$. We emphasize, however, that the game may have other equilibria that are not optimal.

The fundamental problem addressed in this paper is that the state of the world $\theta$ is unknown to the agents and that different agents have different beliefs $\mu_{it}$ on the state of the world. As a consequence, payoffs $u_i(a,\theta)$ cannot be evaluated but estimated and, moreover, estimates of different agents are different. To explain the implications of this latter observation consider the opposite situation in which the agents aggregate their individual beliefs in a common belief $\mu$. In that case, the payoff estimate
\begin{equation}\label{eqn_payoff_global_estimate}
    u_i(\sigma;\mu) := \int_{\theta \in \Theta} u_i(\sigma,\theta)\, d\mu,
\end{equation}
can be evaluated by all agents if we assume that the payoff functions $u_i$ are known globally. Assuming global knowledge of payoffs is not always reasonable and it is desirable to devise mechanisms where agents operate without access to the payoff functions of other agents; see, e.g., \cite{Swenson_et_al_2014}. Still, an important implication of considering the payoffs in \eqref{eqn_payoff_global_estimate} is that optimal behavior is characterized by Nash equilibria. Specifically, for the game defined by the utilities in \eqref{eqn_payoff_global_estimate}, a Nash equilibrium at time $t$ is a strategy profile $\sigma^*$ such that no agent has an interest to deviate unilaterally given the common belief $\mu$. I.e., a strategy $\sigma^* = \{\sigma_i^*, \sigma_{-i}^*\}$ such that for all agents $i$ it holds 
\begin{equation} \label{eqn_nash_equilibrium}
    u_i(\sigma_i^*, \sigma_{-i}^*;\mu) 
        \ \geq\ u_i(\sigma_i, \sigma_{-i}^*;\mu),
\end{equation}
for all other possible strategies $\sigma_i$. Given the existence of the potential function $u$ as stated in \eqref{potential_game_definition}, the Nash equilibrium in \eqref{eqn_nash_equilibrium} with the aggregate belief $\mu$ is a proxy for the maximization of the global payoff $u(a;\mu) := \int_{\theta \in \Theta} u(a,\theta)\, d\mu$. In that sense, it represents the best action that the agents could collectively take if they all had access to common information. For future reference we use $\Gamma(\mu)$ to represent the game with players $\ccalN$, action space $\ccalA$ and payoffs $u_i(a;\mu)$, 
\begin{equation} \label{eqn_game_of_complete_information}
   \Gamma(\mu):= \left\{\ccalN,\ccalA^n,u_i(a;\mu)\right\}.
\end{equation}
The game $\Gamma(\mu)$ is said to have complete information.

When agents have different beliefs, the equilibrium strategies of \eqref{eqn_nash_equilibrium} cannot be used as a target behavior because agents lack the ability to determine if a strategy $\sigma_i$ that they may choose satisfies \eqref{eqn_nash_equilibrium} or not. Indeed, while the complete information game serves as an omniscient reference, agents can only evaluate their expected payoffs with respect to their local beliefs $\mu_{it}$,
\begin{equation}\label{eqn_payoff_local_estimate}
    u_i(\sigma;\mu_{it}) 
          = u_i(\sigma_i, \sigma_{-i};\mu_{it}) 
          = \int_{\theta \in \Theta} u_i(\sigma_i, \sigma_{-i},\theta)\, d\mu_{it}.
\end{equation}
Comparing \eqref{eqn_payoff_global_estimate} and \eqref{eqn_payoff_local_estimate} we see that the fundamental problem of having different beliefs $\mu_{it}$ at different agents is that $i$ lacks information needed to evaluate the expected payoff $u_j(\sigma;\mu_{jt})$ of agent $j$ and, for that reason, the game is said to have incomplete information. A way to circumvent this lack of information is for agent $i$ to keep a belief $\nu^{i}_{jt}$ on the strategy profile of player $j$. If we group these beliefs to define the joint belief $\nu^i_{-it}:=\{\nu^{i}_{jt}\}_{j\neq i}$ that agent $i$ has on the actions of others, it follows that agent $i$ can evaluate the payoff he can expect of different strategies $\sigma_i$ as
\begin{equation}\label{eqn_payoff_with_belief_on_strategies}
    u_i(\sigma_i, \nu^{i}_{-it}; \mu_{it}) 
          = \int_{\theta \in \Theta} u_i(\sigma_i, \nu^i_{-it},\theta)\, d\mu_{it}.
\end{equation}
It is then natural to suggest that agent $i$ should choose the strategy $\sigma_i$ that maximizes the expected payoff in \eqref{eqn_payoff_with_belief_on_strategies}. Such strategy can always be chosen to be an individual action that is termed the best response to the beliefs $\nu^i_{-it}$ on the actions of others and the belief $\mu_{it}$ on the state of the world,
\begin{equation}\label{eqn_best_response}
    a_{it} \in \argmax_{a_i\in\ccalA} u_i(a_i, \nu^i_{-it}; \mu_{it}).
\end{equation}
For future reference we also define the corresponding best expected utility that agent $i$ expects to obtain at time $t$ by playing the best response action in \eqref{eqn_best_response},
\begin{equation} \label{best_response_expected_utility}
    v_i(\nu_{-it}^i; \mu_{it}) := \max_{a_i \in \ccalA} u_i(a_i, \nu_{-it}^i;\mu_{it}).
\end{equation}
We emphasize that $v_i(\nu_{-it}^i; \mu_{it})$ is {\it not} the utility actually attained by agent $i$ at time $t$. That utility depends on the actual state of the world and the actions actually taken by others and is explicitly given by $u_i(a_{it}, a_{-it},\theta)$.

In this paper we consider agents that select best response actions as in \eqref{eqn_best_response} and focus on designing decentralized mechanisms to construct the beliefs $\nu^i_{jt}$ on the actions of others so that the actions $a_{it}$ attain desirable properties. 

In particular, we assume that there is an underlying state learning process so that the local beliefs $\mu_{it}$ on the state of the world converge to a common belief $\mu$ in terms of total variation. I.e., we suppose that
\begin{equation}\label{eqn_state_belief_convergence}
   \lim_{t\to\infty} \TV(\mu_{it},\mu) = 0 \quad \forall i \in \ccalN,
\end{equation}
where the total variation distance $\TV(\mu_{it},\mu):= \sup_{B\in\ccalB(\Theta)} |\mu_{it}(B) - \mu(B)|$ between distributions $\mu_{it}$ and $\mu$ is defined as the maximum absolute difference between the respective probabilities assigned to elements $B$ of the Borel set $\ccalB(\Theta)$ of the space $\Theta$.

The desirable property that we ask of the process that builds the beliefs $\nu^i_{jt}$ on the actions of others is that the actions $a_{it}$ approach one of the Nash equilibrium strategies defined by \eqref{eqn_nash_equilibrium} as the distributions $\mu_{it}$ converge to the common distribution $\mu$. The learning process that we propose for the beliefs $\nu^i_{jt}$ is based on building empirical histograms of past plays as we explain in sections \ref{distributed_fictitious_play_section} and \ref{histogram_sharing_section}. We will show in sections \ref{sec_convergence} and \ref{sec_convergence_histo_sharing} that this procedure yields best responses that approach a Nash equilibrium as long as the convergence of $\mu_{it}$ to $\mu$ is sufficiently fast. We pursue this developments after a pertinent remark.

%
\begin{remark}\label{rmk_bne}\normalfont
The game of incomplete information defined by the payoffs in \eqref{eqn_payoff_local_estimate} has equilibria that do not necessarily coincide with the equilibria of the complete information game $\Gamma(\mu)$. It is easy to think that the best response $a_{it}$ in \eqref{eqn_best_response} yields the best possible utility $u_i$ for agent $i$. In fact, it is possible for agent $i$ to do better by reasoning that other agents are also playing best response to their beliefs. Strategies that yield an equilibrium point of this strategic reasoning are defined as the Bayesian Nash equilibria of the incomplete information game -- see \cite{EksinEtal13_b,Dekel_et_al_2004} for a formal definition. We utilize the best responses in \eqref{eqn_best_response} because determining Bayesian Nash equilibria requires global knowledge of payoffs and information structures. 
\end{remark}

%
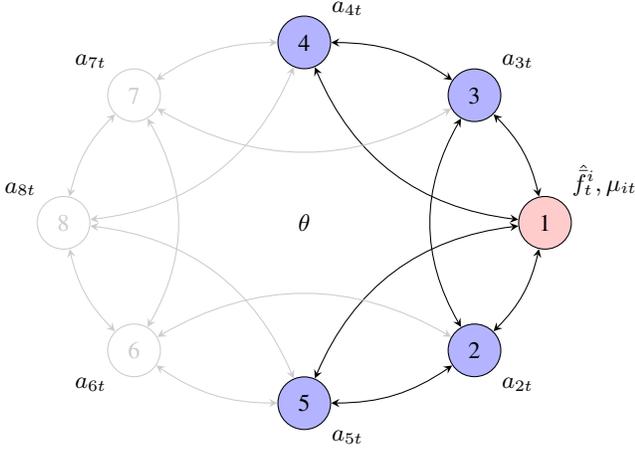
\begin{figure}
\centering
\usetikzlibrary{matrix,arrows,decorations.pathmorphing}
\usetikzlibrary{arrows,automata}

{\small
\tikzstyle{terminal} = [circle, draw=black, inner sep=0pt, minimum size=0.7cm]
\tikzstyle{theta} = [circle, draw=white, inner sep=0pt, minimum size=0.7cm]
\tikzstyle{arrow} = [stealth-stealth, thin]

\begin{tikzpicture}[x=0.8cm, y=0.6cm]
\draw    (   0:0)  node (theta 1)    [theta, fill= white,     label=center:{$\theta$}] {}
        +(-180:4)  node (terminal 8) [terminal, black!20,     label= 135:{$a_{8t}$}] {8} 
        +( 135:4)  node (terminal 7) [terminal, black!20,     label= 135:{$a_{7t}$}] {7}
        +(-135:4)  node (terminal 6) [terminal, black!20,     label=-135:{$a_{6t}$}] {6}
        +(  90:4)  node (terminal 4) [terminal, fill=blue!30, label=  45:{$a_{4t}$}] {4}

        +( -90:4)  node (terminal 5) [terminal, fill=blue!30, label=- 45:{$a_{5t}$}] {5}
        +(  45:4)  node (terminal 3) [terminal, fill=blue!30, label=  45:{$a_{3t}$}] {3}
        +( -45:4)  node (terminal 2) [terminal, fill=blue!30, label=- 45:{$a_{2t}$}] {2}
       ++(   0:4)  node (terminal 1) [terminal, fill=red!20,  label=  45:{$\hat{\bar{f}}^i_{t}, \mu_{it}$}] {1} ;

\draw [arrow, black!20] (terminal 4)  to [bend right  = 17] (terminal 7);
\draw [arrow, black!20] (terminal 7)  to [bend right  = 16] (terminal 8);
\draw [arrow] (terminal 4)  to [bend left  = 17] (terminal 3);
\draw [arrow] (terminal 3)  to [bend left  = 16] (terminal 1);
\draw [arrow, black!20] (terminal 6)  to [bend right = 17] (terminal 5);
\draw [arrow, black!20] (terminal 6)  to [bend left = 16] (terminal 8);
\draw [arrow] (terminal 5)  to [bend right = 17] (terminal 2);
\draw [arrow] (terminal 2) to [bend right = 16] (terminal 1);

\draw [arrow, black!20] (terminal 8)  to [bend right = 30] (terminal 4);
\draw [arrow] (terminal 4)  to [bend right = 30] (terminal 1);
\draw [arrow] (terminal 1) to [bend right = 30] (terminal 5);
\draw [arrow, black!20] (terminal 5)  to [bend right = 30] (terminal 8);
\draw [arrow, black!20] (terminal 7)  to [bend right = 30] (terminal 3);
\draw [arrow] (terminal 3)  to [bend right = 30] (terminal 2);
\draw [arrow, black!20] (terminal 2) to [bend right = 30] (terminal 6);
\draw [arrow, black!20] (terminal 6)  to [bend right = 30] (terminal 7);


\end{tikzpicture}
}
\caption{Distributed fictitious play with observation of neighbors' actions. Agents form beliefs on the strategies of others by keeping a histogram of the average empirical distribution using the observations of actions of their neighbors. E.g., Agent $1$ updates an estimate $\hat{\bar{f}}^i_{t}$ of the average empirical distribution by observing the actions $a_{2t-1},a_{3t-1}, a_{4t-1}, a_{5t-1}$ played by agents 2, 3, 4, and 5 at time $t-1$ [cf. \eqref{local_centroid_empirical_distribution_recursion_2}]. It then selects the best response in \eqref{eqn_best_response} which assumes all other agents are playing with respect to the mixed strategy $\nu_{jt}^i =\hat{\bar{f}}^i_{t}$ given its belief $\mu_{it}$ on the state $\theta$.}
\label{dfp_network}
\end{figure}

%
\section{Distributed fictitious play in symmetric potential games} \label{distributed_fictitious_play_section}

We begin by considering the particular case of symmetric potential games to illustrate concepts, methods, and proof techniques. In a symmetric game, agents' payoffs are permutation invariant in that we have  $u_i(a_i,a_j, a_{-i\setminus j},\theta) = u_j(a_j,a_i, a_{-j \setminus i},\theta)$ for all pairs of agents $i$ and $j$. It follows from this assumption that the game admits at least one consensus Nash equilibrium strategy and that, as a consequence, we  can utilize a variation of fictitious play \cite{Brown_1951,Monderer_Shapley_1996a} in which agents form beliefs on the actions of others by keeping a histogram of actions they have seen taken by other agents in past plays. 

Formally, let $f_{it}\in \reals^{m \times 1}$ denote the empirical histogram of actions taken by $i$ until time $t$ and define the vector indicator function $\Psi(a_{it}) = [\Psi_1(a_{it}),\ldots,\Psi_m(a_{it})]: \ccalA \to \{0,1\}^m$  such that the $k$th component is $\Psi_k(a_{it}) = 1$ if and only if $a_{it}=k$ and $\Psi_k(a_{it}) = 0$ otherwise. Given the definition of the vector indicator function $\Psi(a_{it})$ it follows that the empirical distribution $f_{it}$ of actions taken by $i$ up until time $t>1$ is 
\begin{equation} \label{empirical_distribution}
    f_{it} := \frac{1}{t-1}\sum_{s=1}^{t-1} \Psi(a_{is}).
\end{equation}
The expression in \eqref{empirical_distribution} is simply a vector arrangement of the average number of times that each of the $m$ possible plays $k\in\{1,\dots, m\}$ has been chosen by $i$. Since the game, being symmetric, admits at least one symmetric Nash equilibrium, the histogram of the empirical play of the population as a whole is also of interest. Using the definition of the vector indicator $\Psi(a_{it})$ this empirical distribution is written as
\begin{equation} \label{centroid_empirical_distribution}
   \bar{f}_{t} := \frac{1}{n} \sum_{i=1}^n  
                  \bigg[\frac{1}{t-1}\sum_{s=1}^{t-1} \Psi(a_{is})\bigg]\,.
\end{equation}
In conventional fictitious play, agents play best responses to the composite empirical distribution in \eqref{centroid_empirical_distribution}. Here, however, we assume that agents are part of a connected network $G$ with node set $\ccalN$ and edge set $\ccalE$. Agent $i$ can only interact with neighboring agents $j\in\ccalN_i:=\{j\in\ccalN:(j,i)\in\ccalE\}$. At time $t$, the actions of neighboring agents $j\in\ccalN_i$ become known to agent $i$ either through explicit communication or implicit observation. In this setting, agent $i$ cannot keep track of the empirical distribution in \eqref{centroid_empirical_distribution} because it only observes its neighbors' actions $a_{\ccalN_{i}t} :=\{a_{j t}: j \in \ccalN_i\}$ -- see Fig. \ref{dfp_network}. 

What is possible for agent $i$ to compute is an estimate of \eqref{centroid_empirical_distribution} utilizing the information it has available. This estimate is built by averaging the plays of neighbors so that if we write $i$'s estimate of $\bar{f}_{t}$ as $\hat{\bar{f}}_{t}^i$ it follows that for $t\geq 2$ it holds
\begin{equation} \label{local_centroid_empirical_distribution_recursion}
   \hat{\bar{f}}_{t}^i = \frac{1}{| \ccalN_i|} \sum_{j \in \ccalN_i} 
                         \bigg[\frac{1}{t-1}\sum_{s=1}^{t-1} \Psi(a_{js})\bigg].
\end{equation}
In the distributed fictitious play algorithm considered here, agent $i$ has access to the state belief $\mu_{it}$ and the estimate on the average empirical distribution $\hat{\bar{f}}_{t}^i$ in \eqref{local_centroid_empirical_distribution_recursion}. Agent $i$ proceeds to select the best response action $a_{it}$ that maximizes its expected payoff [cf. \eqref{eqn_best_response}] assuming that all other agents play with respect to the estimated average empirical distribution. 
I.e., the action played by agent $i$ is computed as per \eqref{eqn_best_response} with $\nu_{jt}^i = \hat{\bar{f}}^i_{t}$ for all $j \neq i$. 

Observe that computation of the histograms in \eqref{empirical_distribution} - \eqref{local_centroid_empirical_distribution_recursion} does not require keeping the history of past plays $a_{is}$ for $s<t$. Indeed, the empirical distribution $f_{it}$ in \eqref{empirical_distribution} can be expressed recursively as 
\begin{equation} \label{empirical_distribution_recursion}
    f_{it+1} = f_{it} \ +\  \frac{1}{t}\Big( \Psi(a_{it}) - f_{it}\Big),
\end{equation}
Likewise, we can also write the population's empirical distribution $\bar{f}_{t}$ in \eqref{centroid_empirical_distribution} recursively as
\begin{equation} \label{centroid_empirical_distribution_recursion}
\bar{f}_{t+1}= \bar{f}_{t} + \frac{1}{t}\bigg[\frac{1}{n}\sum_{j=1}^n {\Psi}(a_{j t}) - \bar{f}_{t}\bigg],
\end{equation}
and the same rearrangement permits writing the estimate $\hat{\bar{f}}_{t}^i$ that $i$ keeps of the population's empirical distribution as the recursion
\begin{equation} \label{local_centroid_empirical_distribution_recursion_2}
    \hat{\bar{f}}_{t+1}^i  
        = \hat{\bar{f}}_{t}^i 
          + \frac{1}{t} \bigg[\frac{1}{|\ccalN_i|} 
                        \sum_{j \in \ccalN_i} \Psi(a_{j t})- \hat{\bar{f}}_{t}^i \bigg].
\end{equation}
In all three cases the recursions are valid for $t\geq1$ and we have to make $f_{i1} = \bar{f}_{1} = \hat{\bar{f}}^i_{1} = \bbzero$ for the recursions in \eqref{empirical_distribution_recursion} - \eqref{local_centroid_empirical_distribution_recursion_2} to be equivalent to \eqref{empirical_distribution} - \eqref{local_centroid_empirical_distribution_recursion}. However, subsequent convergence analyses rely on \eqref{empirical_distribution_recursion} - \eqref{local_centroid_empirical_distribution_recursion_2} and allow for arbitrary initial distributions $\hat{\bar{f}}^i_{1}$. This is important because, in general, agent $i$ may have side information on the actions $a_{j1}$ that other agents are to chose at time $t=1$.

We can now summarize the behavior of agent $i$ as follows: (i) At time $t$ update the state's belief to $\mu_{it}$. (ii) Play the best response action $a_{it}$ in \eqref{eqn_best_response} with $\nu_{jt}^i = \hat{\bar{f}}^i_{t}$ for all $j \neq i$, (iii) Learn the actions $a_{it}$ of neighbors, either through observation or communication, and update $\hat{\bar{f}}_{t+1}^i$ as per \eqref{local_centroid_empirical_distribution_recursion_2} -- with the empirical histogram $\hat{\bar{f}}_{1}^i$ initialized to an arbitrary distribution. We show in the following that when all agents follow this behavior, their strategies converge to a consensus Nash equilibrium of the symmetric potential game $\Gamma(\mu)$ [cf. \eqref{eqn_game_of_complete_information}].

%
\subsection{Convergence}\label{sec_convergence}

Game equilibria are not unique in general. Consider then the stage game $\Gamma(\mu)$ with a given common belief $\mu$ on the state of the world $\theta$. The the set of Nash equilibria of the game $\Gamma(\mu)$ contains all the strategies that satisfy \eqref{eqn_nash_equilibrium}, 
\begin{align} \label{BNE_set}
   K(\mu) = \{  \sigma^* &
              : u_i(\sigma^*; \mu) \geq u_i(\sigma_i, \sigma_{-i}^*; \mu),\
              \forall i, \sigma_i
              \}.
\end{align}
In the distributed fictitious play process described by \eqref{eqn_best_response} and \eqref{local_centroid_empirical_distribution_recursion_2} agents assume that all other agents select actions from the same distribution. Therefore, it is reasonable to expect convergence not to an arbitrary equilibrium but to a consensus equilibrium in which all agents play the same strategy. Define then the set of consensus equilibria of the game $\Gamma(\mu)$ as the subset of the Nash equilibria set $K(\mu)$ defined in \eqref{BNE_set} for which all agents play according to the same strategy,
\begin{align} \label{consensus_NE_set}
   C(\mu) = \{\sigma^* = \{\sigma_1^*,\ldots,\sigma_n^*\} \in K(\mu): \sigma^*_1 = \ldots = \sigma^*_n\}.
\end{align}
We emphasize that not all potential games admit consensus Nash equilibria, but the symmetric potential games considered in this section do have a nonempty set of consensus Nash equilibria; see e.g., \cite{Swenson_et_al_2014}. 

We prove here that the best response actions in \eqref{eqn_best_response} are eventually drawn from a consensus equilibrium strategy if the local empirical histograms $\hat{\bar{f}}_{t+1}^i$ are updated according to \eqref{local_centroid_empirical_distribution_recursion_2} and the local state beliefs $\mu_{it}$ converge to the common belief $\mu$ in the sense stated in \eqref{eqn_state_belief_convergence}. In the proof of this result we make use of the following assumptions on the network topology and the state learning process. 

%
\begin{assumption} \label{connectivity_assumption}
The network $G(\ccalN,\ccalE)$ is strongly connected. 
\end{assumption}

%
\begin{assumption} \label{state_learning_assumption}
For all agents $i \in \ccalN$, the local beliefs $\mu_{it}$ converge to a common belief $\mu$ at a rate faster than $\log t/t$,
\begin{equation}
\TV\left(\mu_{it},\mu\right) = O\left(\frac{\log t}{t}\right) .
\end{equation}
\end{assumption}

%
Assumption \ref{connectivity_assumption} is a simple connectivity requirement to ensure that the actions taken by any node eventually become known to all other agents. Assumption \ref{state_learning_assumption} requires that agents reach the common belief $\mu$ fast enough. This assumption is fundamental to subsequent proofs but is not difficult to satisfy -- see Remark \ref{assumption_remark}. We note that the common belief $\mu$ is an \emph{arbitrary} belief on the state $\theta$ to which all agents converge but is not necessarily the optimal Bayesian aggregate of the information that different agents acquire about the state of the world. Validity of these two assumptions guarantees convergence of the best response actions in \eqref{eqn_best_response} as we formally state next.

%
\begin{theorem}\label{action_sharing_convergence}
Consider a symmetric potential game $\Gamma(\mu)$ and the distributed fictitious play updates where at each stage agents best respond as in \eqref{eqn_best_response} with local beliefs $\nu_{jt}^i = \hat{\bar{f}}^i_t$ for all $j\neq i$ formed using \eqref{local_centroid_empirical_distribution_recursion_2} and belief $\mu_{it}$. If Assumptions \ref{connectivity_assumption} and \ref{state_learning_assumption} are satisfied and the initial estimated average beliefs are the same for all agents, i.e., if $\hat{\bar{f}}^i_{1} = \hat{\bar{f}}^j_{1}$ for all $i \in \ccalN$ and $j \in \ccalN$, then the average empirical distribution $\bar{f}_{t}\in \bigtriangleup(\ccalA)$ converges to a strategy that is an element $\sigma_i^*\in \bigtriangleup(\ccalA)$ of a strategy profile $\sigma^*\in \bigtriangleup^n(\ccalA)$ that belongs to the set of consensus Nash equilibria $C(\mu)$ of the symmetric potential game $\Gamma(\mu)$. I.e.,
\begin{align} \label{eqn_average_empirical_convergence}
    \lim_{t\to\infty} \| \bar{f}_{t} - \sigma_i^*\| = 0
\end{align}
where $\sigma_i^*\in \sigma^*$, $\sigma^* \in C(\mu)$, and $\|\cdot\|$ denotes the $\ccalL^2$ norm on the Euclidean space. \end{theorem}
%
%
\begin{myproof} See Appendix \ref{proof_action_sharing_convergence}. \end{myproof}

%
Since in a consensus Nash equilibrium all agents play according to the same strategy, $\sigma_i^* = \sigma_j^*$ for all $i$ and $j$, Theorem  \ref{action_sharing_convergence} also means that the $n$-tuple of the population's average empirical distribution $\bar{f}_t$ is a consensus Nash equilibrium strategy profile $\sigma^*\in C(\mu)$. Notice that this result is not equivalent to showing that each agent's empirical frequency $f_{it}$ is a consensus Nash equilibrium strategy. However, $i$'s model of other agents $\hat{\bar{f}}^i_t$ converges to the average empirical distribution $\bar{f}_t$. In particular, we have $\|\hat{\bar{f}}^i_t - \bar{f}_t \| = O(\log t/t)$ by Lemma \ref{convergence_actions_beliefs}. Hence, agents do learn to best respond to the consensus equilibrium strategy $\bar{f}_t$. In order for agent $i$'s individual empirical frequency $f_{it}$ to converge to the consensus Nash equilibrium strategy $\bar{f}_t$, the utility function should be such that agents are not indifferent between two actions when they best respond in \eqref{eqn_best_response} to the equilibrium strategies of others as we show next.


%
\begin{corollary}\label{consensus_nash_convergence}
In a distributed fictitious play with action sharing, if the potential function $u(\cdot)$ is such that for $\nu^i_{jt} = \bar{f}_t$ for all $j\neq i$ the maximizing action is unique asymptotically, that is, there exists $a^* \in \ccalA$ such that
\begin{equation}\label{eqn_distinct_maximizer}
\lim_{t\to\infty} u(a^*,\nu^i_{-it};\mu) - u(a, \nu^i_{-it};\mu) \geq \epsilon 
\end{equation}
for $\epsilon>0$ and for all $a \in \ccalA \setminus a^*$ then each agent learns to play according to an empirical frequency that is in equilibrium with others' empirical frequencies for any symmetric potential game $\Gamma(\mu)$, that is,
\begin{equation} \label{eqn_convergence_empirical_frequency}
   \lim_{t\to\infty}  \| f_{it} - \bar{f}_t\| = 0.
\end{equation}
\end{corollary}
%
%
\begin{myproof} {See Appendix \ref{dfp_corollary_proof}. }\end{myproof}

%

The condition in \eqref{eqn_distinct_maximizer} says that there exists a single distinct action $a^*$ that strictly maximizes the expected utility asymptotically when other agents follow $\bar{f}_t$. We obtain the result in \eqref{eqn_convergence_empirical_frequency} by leveraging the fact that asymptotically agents' estimates of the average empirical distribution $\hat{\bar{f}}^i_t$ converge to $\bar{f}_t$ and there exists a finite time after which action $a^*$ will be chosen. The result above implies that agents eventually play according to a consensus Nash equilibrium action. Note that the responses of agents during the distributed fictitious play depend on both the state learning process and the process of agents forming their estimates on the average empirical distribution. The results in this section reveal that these two processes can be designed independently as long as both processes converge at a fast enough rate. We make use of this separation in the next section to design a distributed fictitious play process that converges to an equilibrium strategy of any potential game. 

\begin{remark} \label{assumption_remark} The $\log t/t$ convergence rate in Assumption \ref{state_learning_assumption} is satisfied by various distributed learning methods including averaging, e.g., \cite{Jadbabaie_et_al_2003, Kashyap_et_al_2007}; decentralized estimation, e.g., \cite{Ribeiro_ConsensusI, StankovicStankovic, olfati2007distributed, Olfati_et_al_2006, SoummyaKar_ a,chenSayed}; social learning models, e.g., \cite{Shahrampour_et_al_2014, Jadbabaie_et_al_2013}; and Bayesian learning, e.g.,  \cite{Vives_1997, Gale_Kariv, Djuric_2012, Mueller_2013}. In the way of illustration, averaging models have agent $i$ sharing its previous belief on the state with its neighbors and updating its belief by a weighted averaging of observed distributions that follow the recursion
\begin{align}\label{averaging_state_learning}
   \mu_{it}(\theta) = \sum_{j \in \ccalN_i} w_{ij} \mu_{jt-1}(\theta),
\end{align}
for some set of doubly stochastic weights $w_{ij}$. Convergence to a common distribution follows an exponential rate $O(c^t)$ if all the information available to agents are private observations at time $t=1$ \cite{Kashyap_et_al_2007, Nedic_et_al_2009}. In Bayesian learning we can do away with communication altogether and assume that agents keep acquiring private information on $\theta$ that they incorporate in the local beliefs $\mu_{it}$ using Bayes' law. If the local signals are informative, all agents converge to an atomic belief with all probability in the true state of the world. Linear -- meaning $O(1/t)$ -- rates of convergence can be achieved with mild assumptions on the rate of novel information \cite{Vives_1997}. Bayesian updates utilizing neighbors' beliefs are also possible, if computationally cumbersome, and also achieves $O(1/t)$ convergence with mild assumptions \cite{Gale_Kariv,Djuric_2012,Mueller_2013}.
\end{remark}

%
\section{Distributed Fictitious Play in Generic Potential Games}\label{histogram_sharing_section}


For generic potential games we consider a distributed fictitious play process in which agents communicate the histograms they keep on the other agents with their neighbors. I.e., Agent $i$ shares its entire belief $\nu_{-it}^i$ with its neighbors at each time step in addition to its action $a_{it}$. When compared to the distributed fictitious play with action observations, the additional information communicated allows agents to keep distinct beliefs on other agents as we explain in the following. 

Agent $i$ can keep track of the individual empirical histograms of its neighbors $\{f_{jt}\}_{j \in \ccalN_i}$ by \eqref{empirical_distribution_recursion} using observations of the actions of its neighbors $\{a_{jt}\}_{j \in \ccalN_i}$, that is, $\nu^i_{jt+1} = f_{jt+1}$ for $j \in \ccalN_i$. Otherwise, agent $i$ can keep an estimate of the empirical histogram of its non-neighbors $j \notin \ccalN_i$ by averaging the estimates of its neighbors on the non-neighboring agent $j$ $\{\nu_{jt}^k\}_{k \in \ccalN_i}$. I.e. the estimate of agent $i$ on $j \notin \ccalN_i$ is given by 
\begin{equation} \label{belief_update_non_neighbors}
\nu^i_{jt+1} = \sum_{k\in \ccalN}   w^i_{jk} \nu^k_{j t} 
\end{equation}
where $w^i_{jk} > 0$ if and only if $k \in\ccalN_i$ and $\sum_{k\in\ccalN} w^i_{jk}= 1$. Note that in this belief formation, agent $i$ keeps a separate belief on each individual and has the correct estimate of the empirical frequency of its neighbors.

Besides the difference in belief updates the distributed fictitious play is identical to the behavior described in Section \ref{distributed_fictitious_play_section}. To summarize, at time $t$ agent $i$ updates its belief on the state $\mu_{it}$, plays with respect to the best response action $a_{it}$ in \eqref{eqn_best_response} with beliefs $\nu_{-it}^i$, observe actions and beliefs of neighbors $\{a_{jt}, \nu_{-jt}^j\}_{j \in \ccalN_i}$, and update $\nu_{jt+1}^i$ for $j\neq i$ by \eqref{empirical_distribution_recursion} if $j \in \ccalN_i$ or by \eqref{belief_update_non_neighbors} if $j \notin \ccalN_i$. In the following, we show the convergence of the empirical frequencies to a Nash equilibrium strategy for any potential game when agents follow the behavior described above. 

%
\subsection{Convergence}\label{sec_convergence_histo_sharing}

Next, we present the main result of the paper that shows that the best responses in the distributed fictitious play  with histogram sharing converge to a Nash equilibrium strategy \eqref{BNE_set} of any potential game $\Gamma(\mu)$ given the same assumptions on network connectivity and on convergence of the state learning process as in Theorem \ref{action_sharing_convergence}.


%
\begin{theorem} \label{histogram_sharing_convergence}
Consider a potential game $\Gamma(\mu)$ and the distributed fictitious play updates where at each stage agents best respond as in \eqref{eqn_best_response} with local beliefs $\nu_{-it}^i$ formed using \eqref{empirical_distribution_recursion} if $j\in \ccalN_i$ or using \eqref{belief_update_non_neighbors} if $j \notin \ccalN_i$, and state belief $\mu_{it}$. If Assumptions \ref{connectivity_assumption} and \ref{state_learning_assumption} are satisfied then the empirical frequencies of agents $f_t:=\{f_{jt}\}_{j \in \ccalN}\in \bigtriangleup^n(\ccalA)$ defined in \eqref{empirical_distribution} converge to a Nash equilibrium strategy of the potential game with common state of the world beliefs $\mu$, that is, 
\begin{align} \label{histogram_convergence}
 \min_{\sigma^* \in K(\mu)} \| f_t - \sigma^*\| \to 0
\end{align}
where $K(\mu)$ is the set of Nash equilibria of the game $\Gamma(\mu)$ \eqref{BNE_set}. 
\end{theorem}
%
%
\begin{myproof} See Appendix \ref{dfp_histogram_sharing_appendix}. \end{myproof}

%
The above result implies that when agents share their beliefs on others' histograms and based on this information keep an estimate of the empirical distribution of each agent, their responses converge to a Nash equilibrium of the potential game as long as their beliefs on the state reach consensus at a belief $\mu$ fast enough. Theorem \ref{histogram_sharing_convergence} generalizes Theorem \ref{action_sharing_convergence} to the class of potential games given that agents in addition to their actions communicate their beliefs on others with their neighbors. 

%
\begin{figure*}[!t]
\centering
\begin{tabular}{cc} \hspace{-5mm}
\includegraphics[width=0.4\linewidth]{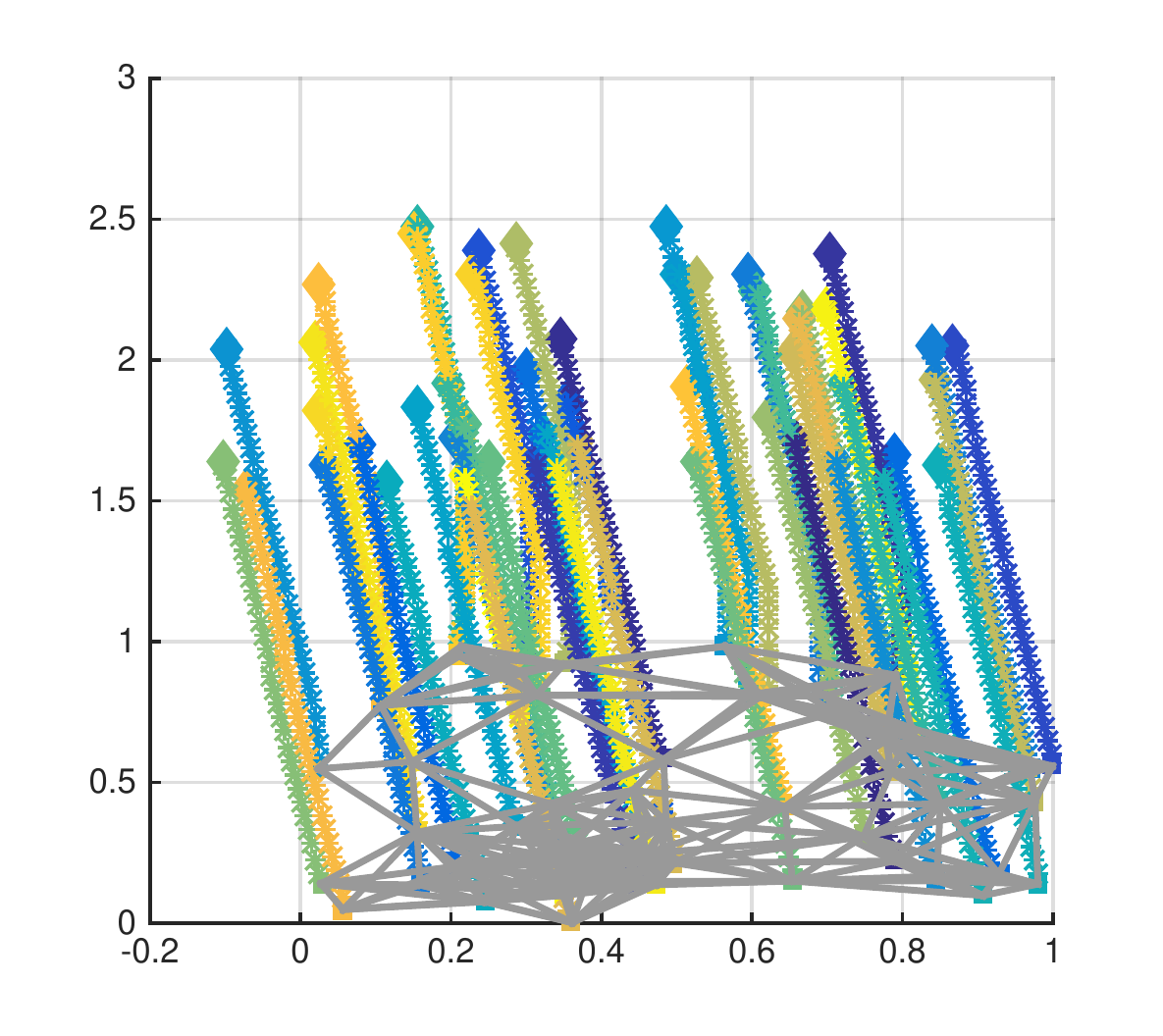}
&\includegraphics[width=0.4\linewidth]{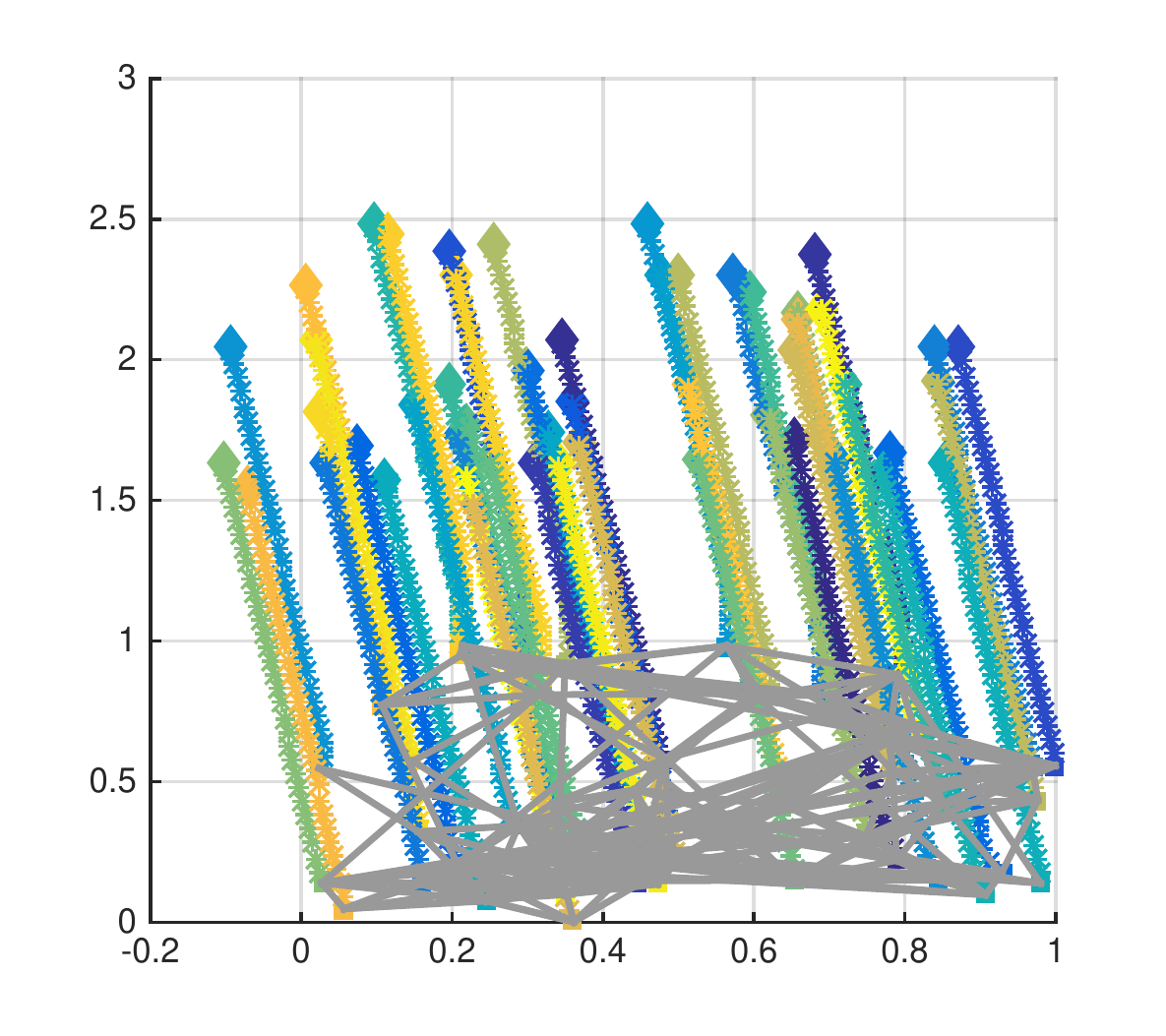}\\
    \fontsize{7}{12}\selectfont (a)
      	       & \fontsize{7}{12}\selectfont (b)
 \end{tabular}
\caption{Position of robots over time for the geometric (a) and small world networks (b). Initial positions and network is illustrated with gray lines. Robots' actions are best responses to their estimates of the state and of the centroid empirical distribution for the payoff in \eqref{beauty_contest_payoff}. Robots recursively compute their estimates of the state by sharing their estimates of $\theta$ with each other and averaging their observations. Their estimates on the centroid empirical distribution is recursively computed using \eqref{local_centroid_empirical_distribution_recursion_2}. Agents align their movement at the direction $95^\circ$ while the target direction is $\theta = 90^\circ$.}
\vspace{-3mm}
\label{positions_geo_small}
\end{figure*}

%
\begin{figure*}[!t]
\centering
\begin{tabular}{cc} \hspace{-5mm}
\includegraphics[width=0.4\linewidth]{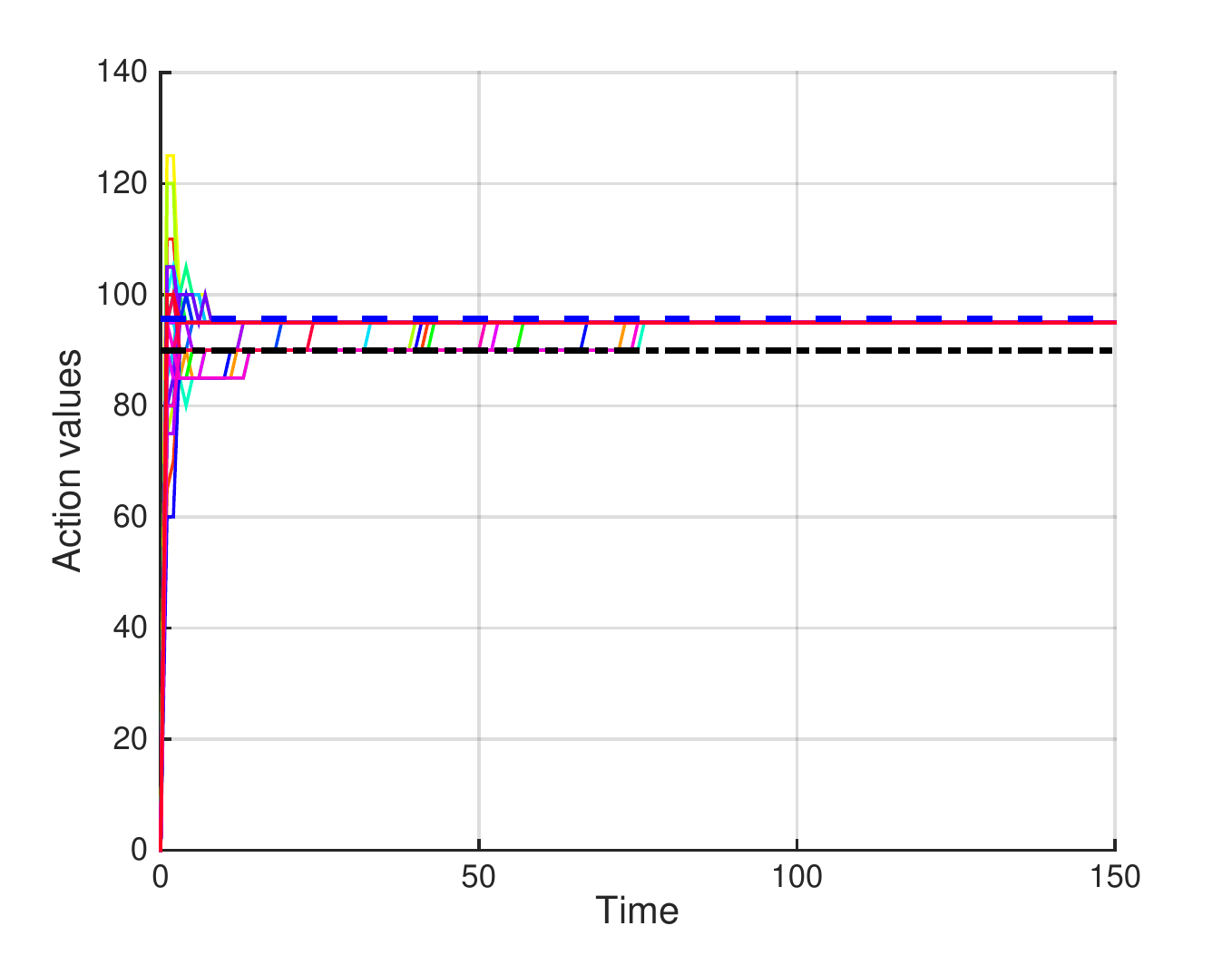}
&\includegraphics[width=0.4\linewidth]{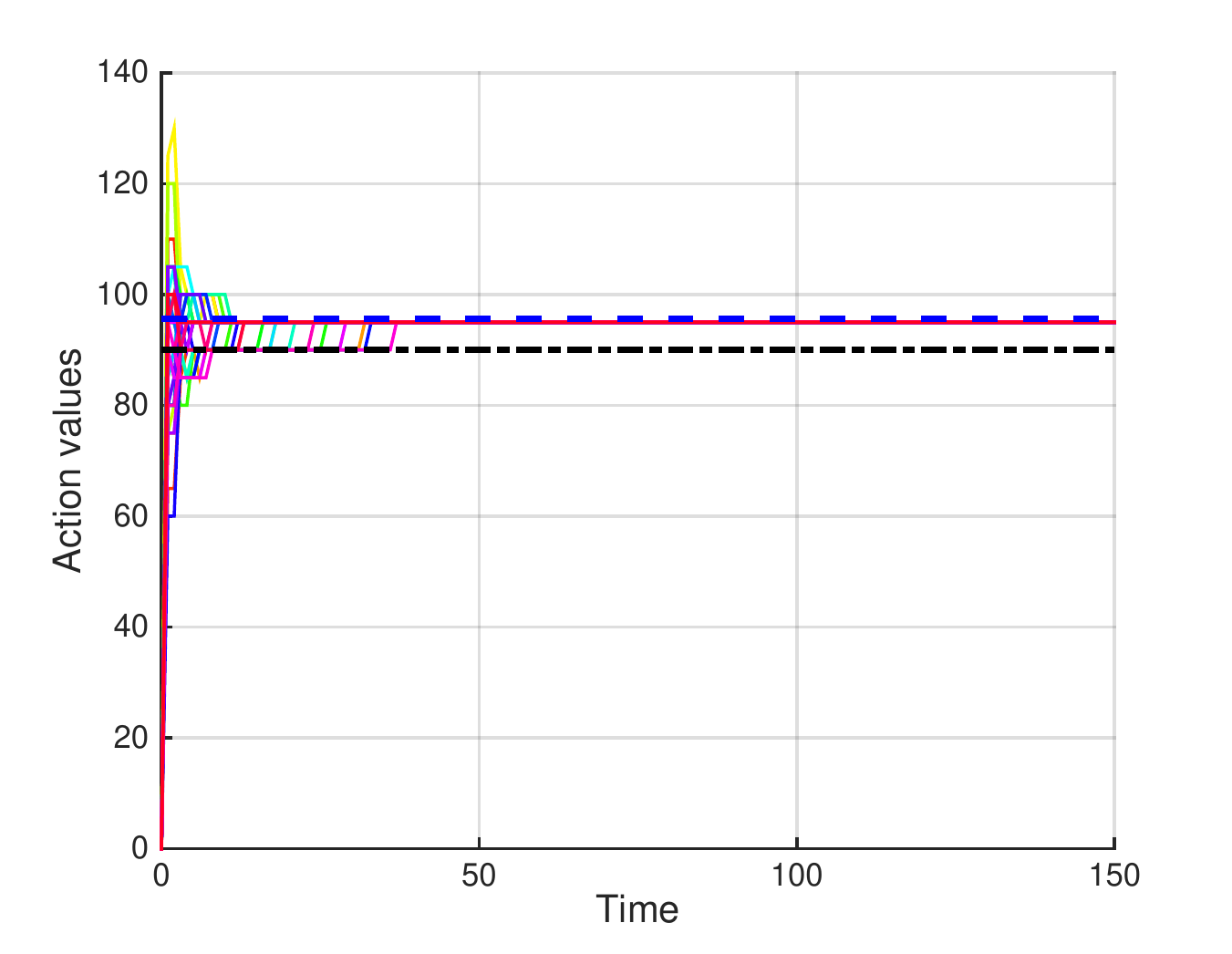}\\
    \fontsize{7}{12}\selectfont (a)
      	       & \fontsize{7}{12}\selectfont (b)
 \end{tabular}
\caption{Distributed fictitious play actions of robots over time for the geometric (a) and small world networks (b). Solid lines correspond to each robots' actions over time. The dotted dashed line is equal to value of the state of the world $\theta= 90^\circ$ and the dashed line is the optimal estimate of the state given all of the signals which is equal to $96.1^\circ$. Agents reach consensus in the movement direction $95^\circ$ faster in the small-world network than the geometric network.
}
\vspace{-2mm}
\label{actions_geo_small}
\end{figure*}

%
\section{Simulations} \label{numerical_examples}

We explore the effects of the network connectivity, the state learning process and the payoff structure on the performance of the algorithm in two games named the beauty contest game, and the target covering game. 

%
\subsection{Beauty contest game}

A network of $n = 50$ autonomous robots want to move in coordination and at the same time follow a target direction $\theta = [0^\circ, 180^\circ]$ in a two dimensional topology. Each robot receives an initial noisy signal related to the target direction $\theta$, 
\begin{equation} \label{signal_i}
\pi_i(\theta) = \theta + \epsilon_i
\end{equation}
where $\epsilon_i$ is drawn from a zero mean normal distribution with standard deviation equal to $20^\circ$. Actions of robots determine their direction of movement and belong to the same space as $\theta$ but are discretized in increments of $5^\circ$, i.e., $\ccalA = (0^\circ, 5^\circ, 10^\circ, \dots, 180^\circ)$. The estimation and coordination payoff of robot $i$ is given by the following utility function
\begin{equation} \label{beauty_contest_payoff}
    u_i(a, \theta) = -    \lambda  (a_i - \theta)^2 
                     - (1-\lambda) \bigg(a_i - \frac{1}{n-1}\sum_{j\neq i}  a_j\bigg)^2
\end{equation}
where $\lambda\in (0,1)$ gauges the relative importance of estimation and coordination. The game is a symmetric potential game and hence admits a consensus equilibrium for any common belief on $\theta$\cite{Eksin_et_al_2013}.

In the following numerical setup, we choose $\theta$ to be equal to $90^\circ$.  We assume that all robots start with a common prior on the centroid empirical distribution in which they believe that each action is drawn with equal probability. They follow the distributed fictitious play updates with action sharing described in Section \ref{distributed_fictitious_play_section}. State learning process is chosen as the averaging model in which robots update their beliefs on the state $\theta$ using \eqref{averaging_state_learning} with initial beliefs formed based on the initial private signal with signal generating function in \eqref{signal_i}. 

\begin{figure*}[!t]
\centering
\begin{tabular}{cc} \hspace{-5mm}
\includegraphics[width=0.48\linewidth]{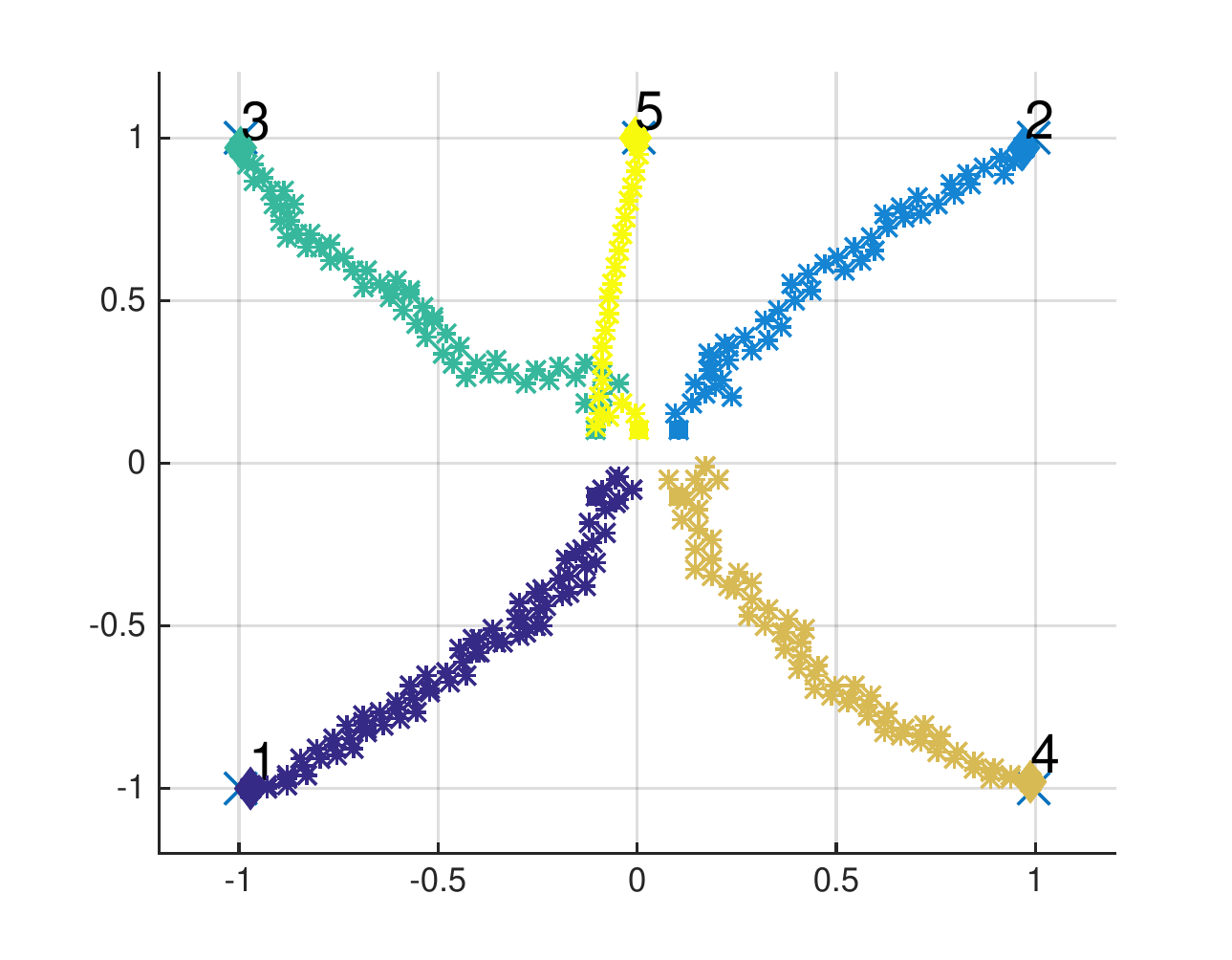}
&\includegraphics[width=0.48\linewidth]{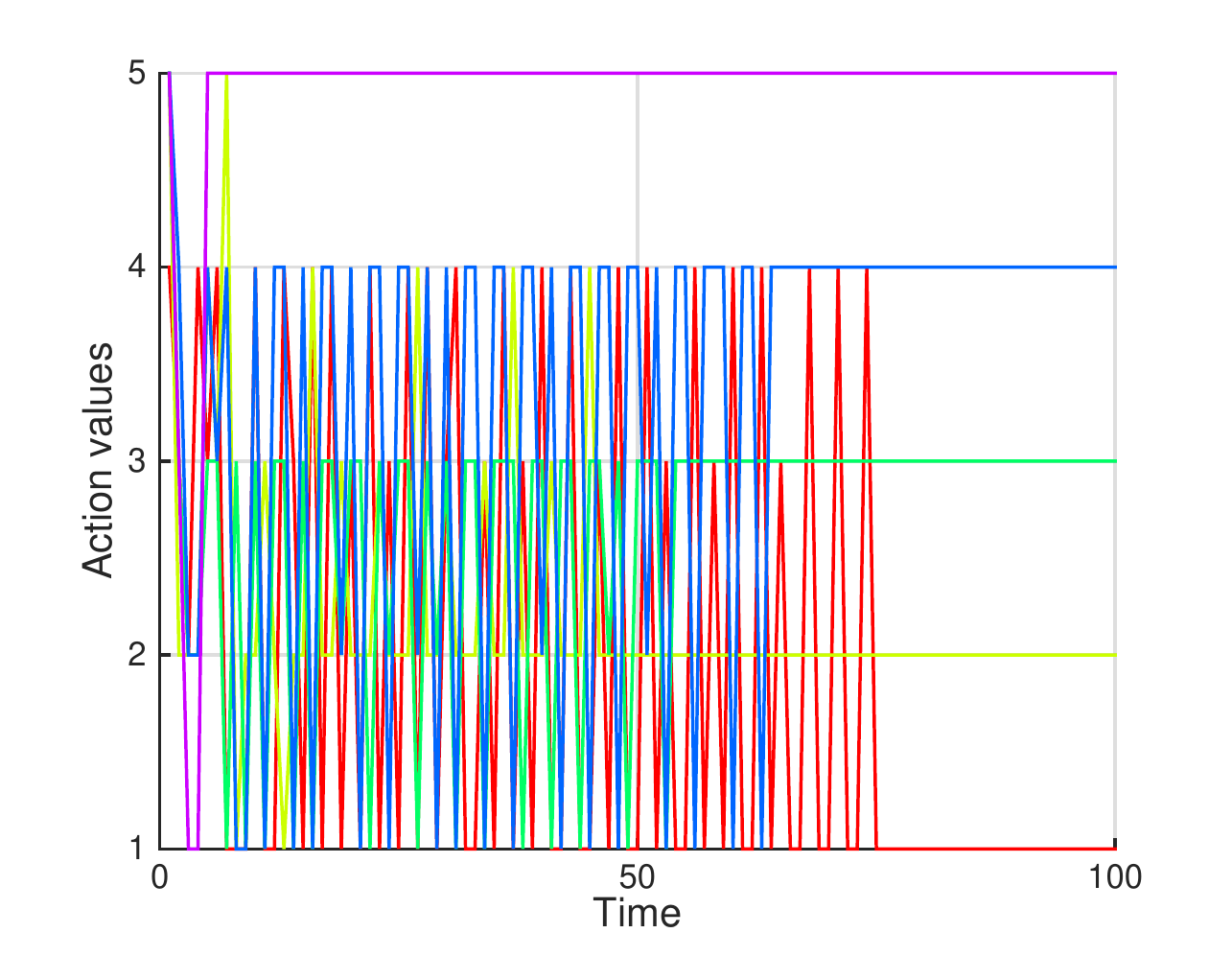}\\
\vspace{-2mm}
    \fontsize{7}{12}\selectfont (a)
      	       & \fontsize{7}{12}\selectfont (b)
 \end{tabular}
\caption{Locations (a) and actions (b) of robots over time for the star network. There are $n=5$ robots and targets.  In (a), the initial positions of the robots are marked with squares. The robots' final positions at the end of 100 steps are marked with a diamond. The positions of the targets are indicated by `$\times$'. Robots follow the histogram sharing distributed fictitious play presented in Section \ref{histogram_sharing_section}. The stars in (a) represent the position of the robots at each step of the algorithm. The solid lines in (b) correspond to the actions of robots over time. All targets are covered by a single robot before 100 steps.}
\vspace{-2mm}
\label{actions_target_assignment}
\end{figure*}

In Figs. \ref{positions_geo_small} and \ref{actions_geo_small}, we plot robots' positions and their actions, respectively. In Fig. \ref{positions_geo_small}, we assume that robot $i$ moves with a displacement of $0.01$ meters in the chosen direction $a_{it}$ at stage $t$. Figs. \ref{positions_geo_small}(a) and \ref{actions_geo_small}(a) correspond to the behavior in a geometric network when robots are placed on a $1$ meter  $\times$ 1 meter square randomly and connecting pairs with distance less than $0.3$ meter between them. Figs. \ref{positions_geo_small}(b) and \ref{actions_geo_small}(b) correspond to the behavior in a small-world network when the edges of the geometric network are rewired with random nodes with probability 0.2. The geometric network illustrated in Fig. \ref{positions_geo_small}(a) has a diameter of $\Delta_g = 5$ with an average length among users equal to 2.5\begin{footnote}{Diameter is the longest shortest path among all pairs of nodes in the network. The average length is the average number of steps along the shortest path for all pairs of nodes in the network.}\end{footnote}. The small world network  illustrated in Fig. \ref{positions_geo_small}(b) has a diameter of $\Delta_r=4$ with an average length among users equal to 2. We observe that the agents reach consensus at the action $95^\circ$ in both networks but the convergence is faster in the small-world network (39 steps) than the geometric network (78 steps). 

We further investigate the effect of the network structure in convergence time by considering 50 realizations of the geometric network and 50 small-world networks generated from the realized geometric networks with rewire probability of 0.2. The average diameter of the realized geometric networks was 5.1 and the average diameter of the realized small-world networks was 4.1. The mean of the average length of the realized geometric networks was 2.27 while the same value was 1.96 for the realized small-world networks. We considered a maximum of 500 iterations for each network. Among 50 realizations of the geometric network, the distributed fictitious play behavior failed to reach consensus in action within 500 steps in 18 realizations while for small-world networks the number of failures was 5. The average time to convergence among the 50 realizations was 228 steps for the geometric network whereas the convergence took 100 steps for the small-world network on average. In addition, convergence time for the small-world network is observed to be shorter than the corresponding geometric network in all of the runs except one.

%
\subsection{Target covering game}

$n$ autonomous robots want to cover $n$ targets. The position of a target $k \in \ccalT:=\{1,\dots,n\}$ on the two dimensional space is denoted by $\theta_k \in\reals^2$ and are not known by the robots. Robot $i$ starts from an initial location $x_i \in \reals^2$ and makes noisy observations $s_{ik0}$ of the location of target $k$ coming from normal distribution with mean $\theta_k$ and standard deviation equal to $\sigma \bbI$ where $\bbI$ is the $2 \times 2$ identity matrix and $\sigma >0$ for all $k \in \ccalT$. At each stage robots choose one of the targets, that is, $\ccalA = \ccalT$ and receives a payoff from covering that target that is inversely proportional to its distance from the target if no other robot is covering it, that is, the payoff of robot $i$ from covering target $k \in(1,\dots, n)$ $a_i = k$  is given by 
\begin{align} \label{target_assignment_payoff}
u_i(a_i = k, a_{-i},\theta) = \bbone\left(\sum_{j \neq i} \bbone\left(a_j = k\right) = 0\right) h(x_i, \theta_k)
\end{align}
where $\bbone(\cdot)$ is the indicator function and $h(\cdot)$ is a reward function inversely proportional to the distance between the target and the robot's initial position $x_i$, e.g., $\|x_i - \theta_k\|^{-2}$. The first term in the multiplication above is one if no one else chooses target $k$ otherwise it is zero. The second term in the multiplication decreases with growing distance between robot $i$'s initial position $x_i$ and the target $k$'s position $\theta_k$. 

When all of the robots start from the same location, that is, $x_i = x$ for all $i\in \ccalN$, the game with payoffs above can be shown to be a potential game
by using the definition of potential games in \eqref{potential_game_definition}. Furthermore, the game is symmetric. 
In this setup, we would like each robot to assign itself to a single target different from the rest of the robots, that is, we are interested in convergence to a pure strategy Nash equilibrium in which each robot picks a single action similar to the target assignment games considered in \cite{Arslan_et_al_2007}. Observe that the target covering game can not have a pure consensus equilibrium strategy.  To see this, assume that all robots cover the same target then they all receive a payoff of zero. Any robot that deviates to another target receives a positive payoff. Therefore, there cannot be a pure consensus strategy equilibrium. As a result, instead of the action sharing scheme, we consider the histogram sharing distributed fictitious play by which it is possible but not guaranteed that the robots converge to a pure strategy Nash equilibrium.

In the numerical setup, we consider $n =5$ robots with the payoffs in \eqref{target_assignment_payoff} and $n$ targets. The locations of targets are respectively given as follows $\theta_1 = (-1,-1)$, $\theta_2 = (1,1)$, $\theta_3 = (-1,1)$, $\theta_4 = (1,-1)$, $\theta_5 = (0,1)$.  We consider the case that the initial positions of robots are different from each other with the reward function $h(x_i,\theta_k) = \|x_i - \theta_k\|^{-2}$. Specifically, the initial positions of the robots equal to $x_1 = (-0.1,-0.1)$, $x_2 = (0.1,0.1)$, $x_3 = (-0.1,0.1)$, $x_4 = (0.1,-0.1)$, and $x_5 = (0,0.1)$. Robots make noisy observations $s_{ikt}$ for all $k\in \ccalT$ after each step. The observations have the same distribution as $s_{ik0}$ with $\sigma = 0.2$ meters. We assume that the robots update their beliefs on the positions of targets using the Bayes' rule based on the observations. 

Figs. \ref{actions_target_assignment}(a)-(b) shows the movement of robots and actions of robots over time, respectively, for the star network. We assume that robots move by a distance of $0.02$ meters along the estimated direction of the target they choose at each step of the distributed fictitious play. The estimated direction is a straight line from the current position to the estimated position of the chosen target. I.e., the robots make observations and decisions in every 0.02 meters. Finally, we assume that the robot covers the target if it is 0.05 meters away from a target and no other robot covers it. In figs. \ref{actions_target_assignment}(a)-(b), we observe that each robot comes to 0.05 meters neighborhood of a target within 100 steps. Furthermore, the robots cover all of the targets, that is, they converge to a pure Nash equilibrium. 

\begin{figure}
\centering
\includegraphics[width=0.9\linewidth]{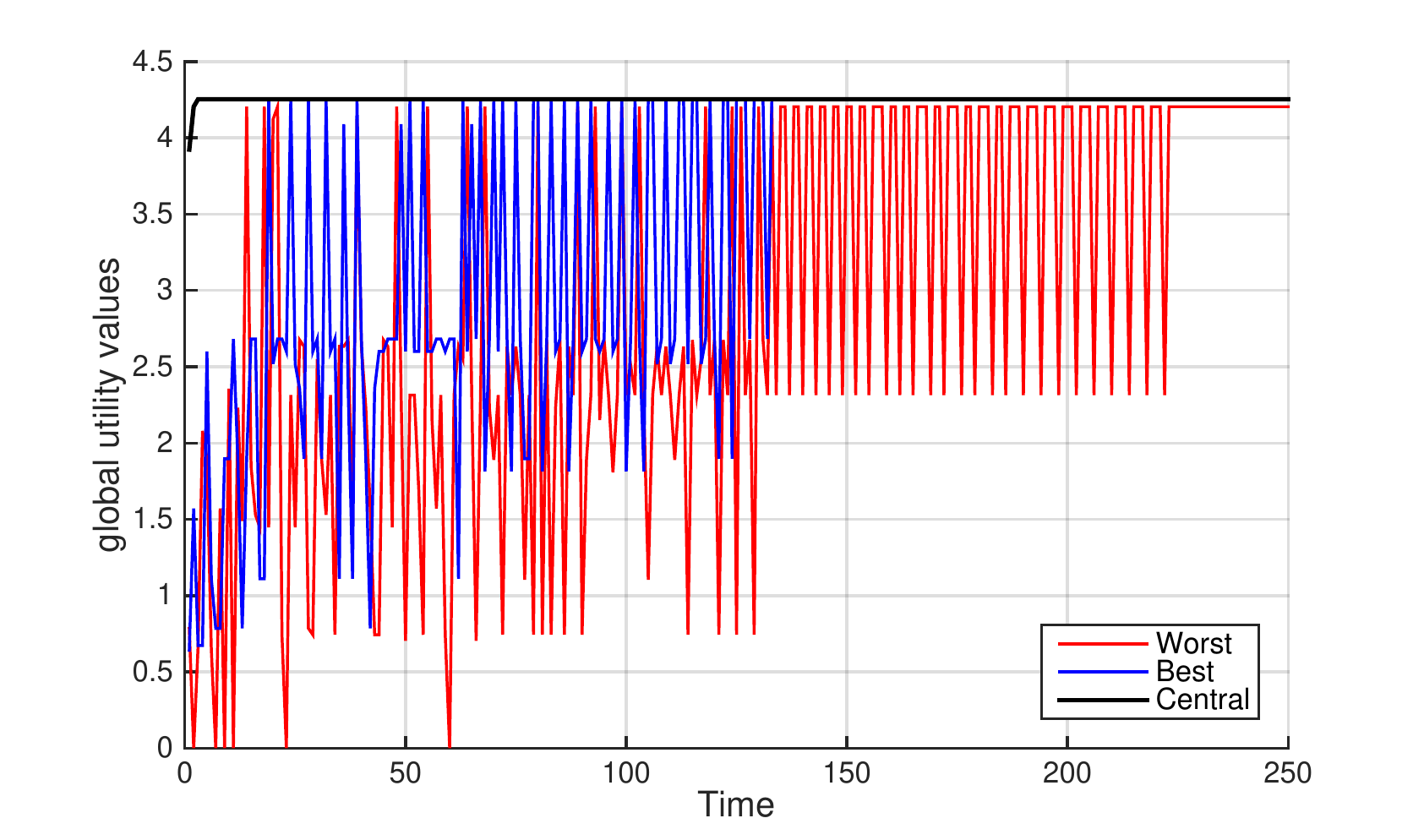}
\caption{Comparison of the distributed fictitious play algorithm with the centralized optimal solution. Best and worst correspond to the runs with the highest and lowest global utility in the distributed fictitious play algorithm.  The algorithm converges to the global optimal point in 40 runs out of a total of 50 runs.}
\label{compare_dfp}
\end{figure}

Next, we compare the distributed fictitious play algorithm to the centralized (optimal) algorithm. In the centralized algorithm, at the beginning of each step agents aggregate their signals and then take the action to maximize the expected global objective defined as the sum of the utilities of all \eqref{target_assignment_payoff}. Since there exists multiple equilibria in the complete information target coverage game, it is not guaranteed that the distributed fictitious play algorithm converges to the optimal equilibrium at each run. For this purpose, we considered 50 runs of the algorithm where in each run signals are generated from different seeds. We assume that the algorithm has converged when each target is covered by a robot within 0.05 meters from the target. In Fig. \ref{compare_dfp}, we plot the evolution of the global utility with respect to time for the distributed fictitious play algorithm runs with the best and the worst final payoff, and for the centralized algorithm. The best final configuration overlaps with the final centralized solution which is given by $a = [1,2,3,4,5]$ resulting in a global objective value of $4.25$. The worst final configuration is given by $a = [1,5,3,4,2]$ resulting in a global objective value of $4.20$. 

%

%
\section{Conclusion}

This paper considered the optimal behavior of multi-agent systems with uncertainty on the state of the world. The fundamental problem of interest was to have a model of optimal agent behavior given a global objective that depends on the state and actions of all the agents when agents have different beliefs on the state. We posed the setup as a potential game with the global objective as the common utility of the multi-agent system and set the optimal behavior as a Nash equilibrium strategy of the game when agents have common beliefs on the state of the environment. We presented a class of distributed algorithms based on the fictitious play algorithm in which agents reach an agreement on their state beliefs asymptotically through an exogenous process, build beliefs on the behavior of others based on information from neighbors and best respond to their expected utility given their beliefs on the state and others. 

We considered two information exchange scenarios for the algorithm where in the first scenario agents communicated their actions. For this scenario we showed that when the agents keep track of the population's average empirical frequency of actions as a belief on the behavior of every other individual, their behavior converges to a consensus Nash equilibrium of any symmetric potential game with common beliefs on the state. In the second scenario we considered agents exchanging their entire beliefs on others in addition to their actions. For this scenario we proposed averaging of the observed histograms as a model for keeping beliefs on the behavior of others and showed that their empirical frequency converges to a Nash equilibrium of any potential game. We exemplified the algorithm in a coordination game -- a symmetric potential game --  and a target covering game -- a potential game. In these examples, we observed that the diameter of the network is influential in convergence rate where the shorter the diameter is, the faster the convergence is.

\begin{appendices}

%


\section{Definitions and intermediate convergence results}

We define notions that relate closeness of a strategy to the set of consensus Nash equilibria of the game $\Gamma(\mu)$.

The distance of a strategy $\sigma \in\bigtriangleup^n(\ccalA)$ from the set of consensus Nash equilibria $C(\mu)$ is given by 
\begin{equation} \label{distance_definition}
d(\sigma, C(\mu)) = \min_{g \in C(\mu)} \| \sigma - g\|.
\end{equation}

The set of consensus strategies that are $\epsilon$ away from the consensus Nash equilibrium set \eqref{consensus_NE_set} is the $\epsilon$-consensus Nash equilibrium strategy set, that is,
\begin{align}
C_\epsilon(\mu) = \{&\sigma^* \in \bigtriangleup^n(\ccalA) : u_i(\sigma^*; \mu) \geq u_i(\sigma_i, \sigma_{-i}^*; \mu)- \epsilon, \nonumber\\
& \forall \sigma_i \in \bigtriangleup(\ccalA), \forall i, \sigma_1=\sigma_2 = \dots = \sigma_n\}
\end{align}
for $\epsilon>0$.

We define the $\delta$-consensus neighborhood of $C(\mu)$ as  
\begin{align}
B_\delta(C(\mu)) = \big\{\sigma \in \bigtriangleup^n(\ccalA)&: d(\sigma, C(\mu)) < \delta,\nonumber\\
& \sigma_1 = \sigma_2= \dots = \sigma_n\big\}.
\end{align} 
Note that the $\delta$ consensus neighborhood is defined as the set of consensus strategies that are close to the set $C(\mu)$. We can similarly define the $\epsilon$ Nash equilibrium set $K_{\epsilon}(\mu)$ and $\delta$ neighborhood of $K(\mu)$ in \eqref{BNE_set} as $B_\delta(K(\mu))$ by removing the agreement constraint on the equilibrium strategies \cite{Swenson_et_al_2014}.

The following intermediate results can be found in Appendix B in \cite{Swenson_et_al_2014}. They are stated here for completeness. 
%
\begin{lemma} \label{L_convergence}
If the processes $g_{t}\in \bigtriangleup^n(\ccalA)$ and $h_t\in \bigtriangleup^n(\ccalA)$ are such that  $\|g_{-it} - h_{-it}\| = O(\log t/t)$ for all $i \in \ccalN$ and the state learning process for all $i \in \ccalN$ generates estimate beliefs $\{\{\mu_{it}\}_{t=0}^\infty\}_{i \in \ccalN}$ that satisfy Assumption \ref{state_learning_assumption}, then for a potential payoff $u$ in \eqref{potential_game_definition} the following is true for the expected utility of best response behavior $v(\cdot)$ in \eqref{best_response_expected_utility},
\begin{equation}
\|v(g_{-it}; \mu_{it}) - v(h_{-i t}; \mu)\|=O(\frac{\log t}{t}).
\end{equation}
\end{lemma}
\begin{myproof}
The proof is detailed in Lemma 4 in \cite{Swenson_et_al_2014}. The proof follows by first making the observation that the expected utility defined in \eqref{eqn_expected_utility} for the potential function is Lipschitz continuous, and second using the definition of the Lipschitz continuity to bound the difference between the best response expected utilities in \eqref{best_response_expected_utility} for $g_{-it}, \mu_{it}$ and $h_{-it}, \mu$ by the distance between $g_{-it},\mu_{it}$ and $h_{-it},\mu$ multiplied by the Lipschitz constant. 
\end{myproof}
%


\begin{lemma} \label{beta_convergence}
If $\sum_{t=1}^T \frac{\alpha_{t}}{t} < \infty$ for all $T >0$, $\|\alpha_{t} - \beta_{t}\| = O(\frac{\log t}{t})$ and $\beta_{t+1}\geq 0$ then $\sum_{t=1}^T \frac{\beta_{t}}{t} < \infty$ as $T \to \infty$. 
\end{lemma}
\begin{myproof}
Refer to the proof of Lemma 5 in \cite{Swenson_et_al_2014}.
\end{myproof}


\begin{lemma} \label{convergence_consensus_equilibrium}
Denote the $n$-tuple of the average empirical distribution with $\bar{f}_t^n:=\{\bar{f}_t, \dots,\bar{f}_t\}$. If for any $\epsilon >0$ the following holds
\begin{equation} \label{number_away_from_epsilon_consensus}
\lim_{T\to \infty} \frac{\#\{1\leq t\leq T: \bar{f}_t^n \notin C_\epsilon(\mu)\}}{T} = 0 
\end{equation}
then $\lim_{t \to \infty} d(\bar{f}_t^n, C(\mu))=0$ where $d(\cdot,\cdot)$ is the distance defined in \eqref{distance_definition}.
\end{lemma}
\begin{myproof}
By Lemma 7 in \cite{Swenson_et_al_2014}, \eqref{number_away_from_epsilon_consensus} implies that for a given $\delta>0$ there exists an $\epsilon >0$ such that
\begin{equation} \label{number_away_from_neighborhood}
\lim_{T\to \infty} \frac{\#\{1\leq t\leq T: \bar{f}_t^n \notin B_\delta(C(\mu))\}}{T} = 0 
\end{equation}
Using above equation, the result follows by Lemma 1 in \cite{Monderer_Shapley}.
\end{myproof}


%
%
%

%
%

\begin{lemma} \label{convergence_equilibrium}
For the potential game with function $u(\cdot)$ in \eqref{potential_game_definition} and expected best response utility $v(\cdot)$ \eqref{best_response_expected_utility}, consider a sequence of distributions $f_t \in\bigtriangleup^n(\ccalA)$ for $t=1,2,\dots$ and a common belief on the state ${\mu} \in \bigtriangleup(\Theta)$. Define the process $\beta_t:= \sum_{i=1}^n v(f_{-it}; \mu) - u(f_{it}, f_{-it}; \mu)$ for $t=1,2,\dots$. If 
\begin{align} \label{beta_process}
\lim_{T\to \infty} \frac{1}{T}\sum_{t=1}^{T} \frac{\beta_{t}}{t}  = 0
\end{align}
then $\lim_{t \to \infty} d( {f}_t, K(\mu))=0$ where $f_t = \{f_{1t},\dots, f_{nt}\}$.
\end{lemma}

\begin{myproof}
By {Lemma 6 in \cite{Swenson_et_al_2014}}, the condition \eqref{beta_process} implies that for all $\epsilon >0$
\begin{equation} \label{number_away_from_epsilon_BNE}
\lim_{T\to \infty} \frac{\#\{1\leq t\leq T: {f}_t \notin K_\epsilon(\mu)\}}{T} = 0. 
\end{equation} 
%

{By Lemma 7 in \cite{Swenson_et_al_2014}}, \eqref{number_away_from_epsilon_BNE} implies that for all $\delta>0$ the following is true
\begin{equation} \label{number_away_from_neighborhood_BNE}
\lim_{T\to \infty} \frac{\#\{1\leq t\leq T: {f}_t \notin B_\delta(K(\mu))\}}{T} = 0 
\end{equation}
The above convergence result yields desired convergence result by Lemma 1 in \cite{Monderer_Shapley}.
\end{myproof}


\section{Proof of Theorem \ref{action_sharing_convergence}} \label{proof_action_sharing_convergence}

Before we prove the theorem, we present an intermediate result that shows the convergence rate of the belief of agent $i$ on the population's average empirical distribution $\hat{\bar{f}}_{t}^i$ to the true average empirical distribution of the population $\bar{f}^i_{t}$.

\begin{lemma} \label{convergence_actions_beliefs} 
Consider the distributed fictitious play in which the centroid empirical distribution of the population $\bar{f}_{t}$ evolves according to \eqref{centroid_empirical_distribution_recursion} and agents update their estimates on the empirical play of the population $\hat{\bar{f}}^i_{t}$ according to \eqref{local_centroid_empirical_distribution_recursion_2}. If the network satisfies Assumption \ref{connectivity_assumption} and the initial beliefs are the same for all agents, i.e., $\hat{\bar{f}}^i_{1} = \bar{f}_{1}$ for all $i \in \ccalN$, then $\hat{\bar{f}}^i_{t}$ converges in norm to $\bar{f}_{t}$ at the rate $O({\log t}/{t})$, that is, $\|\hat{\bar{f}}^i_{t} - \bar{f}_{t}\| = O(\frac{\log t}{t})$
\end{lemma}
\begin{myproof}
See Appendix A in \cite{Swenson_et_al_2014} for a proof.
\end{myproof}

Observe that the above result is true irrespective of the game that the agents are playing and uncertainty in the state. The proof leverages on the fact that the change in the centroid empirical distribution is at most $1/t$ by the recursion in \eqref{centroid_empirical_distribution_recursion}. Then by averaging observed actions of neighbors in a strongly connected network the beliefs of agent $i$ on the centroid empirical distribution evolves faster than the change in the centroid empirical distribution.

\begin{myproof}[Theorem \ref{action_sharing_convergence}]
Given the recursion for the average empirical distribution in \eqref{centroid_empirical_distribution_recursion}, we can write the expected utility for the potential function $u(\cdot)$ when all agents follow the centroid empirical distribution $\bar{f}_{t+1}$ and have identical beliefs $\mu$ as follows
\begin{equation}
u(\bar{f}_{t+1}^n;\, \mu) = u\left(\bar{f}_t^n  + \frac{1}{t}\left(\frac{1}{n} \sum_{i=1}^n \Psi(a_{it}) - \bar{f}_{t}^n\right);\, \mu\right)
\end{equation}
where $\bar{f}_t^n:=\{\bar{f}_t,\dots,\bar{f}_t\}\in \bigtriangleup^n(\ccalA)$ is the $n$-tuple of the average population empirical distribution. Define the average best response strategy at time $t$ $\bar{\Psi}(a_{t}) := \frac{1}{n} \sum_{i=1}^n \Psi(a_{it})$. By the multi-linearity of the expected utility, we expand the above expected utility as follows \cite{Monderer_Shapley}
\begin{align} \label{multilinear_expected_utility}
u(\bar{f}_{t+1}^n;&\, \mu) = u(\bar{f}_{t}^n; \mu)  + \nonumber \\
&\frac{1}{t} \sum_{i =1}^n u(\bar{\Psi}(a_{t}), \bar{f}^{n-1}_t; \mu) - u(\bar{f}_t, \bar{f}^{n-1}_t;\mu) 
+  \frac{\delta}{t^2}
\end{align}
where the first order terms of the expansion are explicitly written and the remaining higher order terms are collected to the term $\delta/t^2$.  

Consider the total utility term in \eqref{multilinear_expected_utility} where agent $i$ is playing with respect to the average best response strategy at time $t+1$ $\bar{\Psi}(a_{t})$ and remaining agents use the average empirical distribution $\bar{f}^{n-1}_t$, $\sum_{i=1}^nu(\bar{\Psi}(a_{t}), \bar{f}^{n-1}_t; \mu)$. By the definition of the average best response strategy, we write the term in consideration as 
\begin{align}
\sum_{i=1}^n &u\left(\bar{\Psi}(a_{t}), \bar{f}^{n-1}_t; \mu\right)  = \sum_{i=1}^n u\left(\frac{1}{n} \sum_{i=1}^n \Psi(a_{it}), \bar{f}^{n-1}_t; \mu\right).
\end{align}
The following equality can be shown by using the multi-linearity of expectation and permutation invariance of the utility \cite{Swenson_et_al_2014},
\begin{align} \label{permutation_invariant_best_response}
\sum_{i=1}^n &u(\bar{\Psi}(a_{t}), \bar{f}^{n-1}_t; \mu)  =  \sum_{i =1}^n u(\Psi(a_{it}), \bar{f}^{n-1}_t; \mu). 
\end{align}
The above equality means that the total expected utility when agents play with the average best response strategy at time $t$ $\bar{\Psi}(a_{t})$ against the average empirical distribution $\bar{f}^{n-1}_t$ at time $t$ is equal to the total expected utility when agents best respond to the average population empirical distribution at time $t$. 

We substitute in the above equality \eqref{permutation_invariant_best_response} for the corresponding term in \eqref{multilinear_expected_utility} to get the following
\begin{align}
u(\bar{f}_{t+1}^n;&\, \mu) = u(\bar{f}_{t}^n; \mu)  + \nonumber \\
&\frac{1}{t} \sum_{i =1}^n u(\Psi(a_{it}), \bar{f}^{n-1}_t; \mu) - u(\bar{f}_{t},\bar{f}^{n-1}_t;\mu) +  \frac{\delta}{t^2}.
\end{align}
 We can upper bound the right hand side by adding $|\delta|/t^2$ to the left hand side.
\begin{align} 
u(\bar{f}_{t+1}^n;&\, \mu) - u(\bar{f}_{t}^n; \mu)  +  \frac{|\delta|}{t^2} \geq \nonumber\\
&\frac{1}{t} \sum_{i =1}^n u(\Psi(a_{it}), \bar{f}^{n-1}_t; \mu) - u(\bar{f}_{t},\bar{f}^{n-1}_t;\mu) 
\end{align}

Define $L_{i t} := v_i(\nu^i_{-it}; \mu_{i t}) - u(\Psi(a_{it}), \bar{f}^{n-1}_t; \mu)$ where $\nu^i_{jt} = \hat{\bar{f}}^i_t$ for $j \neq i$. Note that since agents have identical payoffs, we can drop the subindex of the expected utility of agent $i$ when it best responds to the strategy profile of others $v_i(\cdot)$ defined in Section \ref{model_section} to write it as $v(\cdot)$. Now we add and subtract $\sum_{i=1}^n L_{i t} / t$ to both sides of the above equation to get the following inequality,
\begin{align}
u(\bar{f}_{t+1}^n;& \, \mu) - u(\bar{f}_{t}^n; \mu) + \frac{|\delta|}{t^2} \nonumber \\
&+ \frac{1}{t} \sum_{i =1}^n  v(\nu^i_{-it}; \mu_{it})- u(\Psi(a_{i t}), \bar{f}_{t}^{n-1}; \mu) \nonumber\\ 
&\geq \frac{1}{t} \sum_{i =1}^n v(\nu^i_{-it}; \mu_{it}) - u(\bar{f}_{t},\bar{f}^{n-1}_t;\mu).
\end{align}
Summing the inequalities above from time $t=1$ to time $t= T$, we get 
\begin{align} \label{upper_bound_alpha}
u(\bar{f}_{T+1}^n;&\, \mu) - u(\bar{f}_{1}^n; \mu)+ \sum_{t=1}^{T+1}\frac{|\delta|}{t^2} + \sum_{t=1}^{T+1} \sum_{i =1}^n \frac{L_{i t}}{t} \nonumber \\ 
&\geq\sum_{t=1}^{T+1} \frac{1}{t} \sum_{i =1}^n v(\nu^i_{-it}; \mu_{i t}) - u(\bar{f}_{t},\bar{f}^{n-1}_t;\mu).
\end{align}
Next we define the following term that corresponds to the inside summation on the right hand side of the above inequality,
\begin{equation} 
\alpha_{t} := \sum_{i =1}^n v(\nu^i_{-it}; \mu_{i t}) - u(\bar{f}_{t}, \bar{f}_{t}^{n-1};\mu). 
\end{equation}
The term $\alpha_t$ captures the total difference between expected utility when agents best respond to their beliefs on the average population empirical distribution $\nu^i_{jt} = \hat{\bar{f}}^i_t$ and their beliefs on $\theta$ $\mu_{it}$, and when they follow the current centroid empirical distribution $\bar{f}_{t}$ with common beliefs on the state $\mu$. 
Note that by Lemma \ref{convergence_actions_beliefs} and Assumption \ref{state_learning_assumption} the conditions of Lemma \ref{L_convergence} are satisfied, that is, $\|v(\nu^i_{-it}; \mu_{it}) - u(\bar{f}_{t}, \bar{f}_{t}^{n-1}; \mu)\|=O(\log t/t)$. By the assumption that utility value is finite and Lemma \ref{L_convergence}, the left hand side of \eqref{upper_bound_alpha} is finite. That is, there exists a $\bar{B} >0$ such that
\begin{equation}
\bar{B} \geq \sum_{t=1}^{T+1} \frac{\alpha_{t}}{t}.
\end{equation}
for all $T>0$. Next, we define the following term
\begin{align}
\beta_{t} := \sum_{i = 1}^n v(\bar{f}_{t}^{n-1}; \mu) - u(\bar{f}_{t}, \bar{f}_{t}^{n-1}; \mu) 
\end{align}
that captures the difference in expected payoffs when agents best respond to the centroid empirical distribution $\bar{f}_{t}^{n-1}$ for others given the common asymptotic belief $\mu$, and when everyone follows the current centroid empirical distribution $\bar{f}_{t}$ with common beliefs on the state $\mu$. When we consider the difference between $\alpha_{t}$ and $\beta_{t}$,  the following equality is true by Lemma \ref{L_convergence},
\begin{align}
\| \alpha_{t} - \beta_{t}\| &= \| \sum_{i = 1}^n v(\nu^i_{-it}; \mu_{i t}) - v(\bar{f}_{t}^{n-1}; \mu) \| = O(\frac{\log t}{t}).
\end{align}

Further $\beta_{t} \geq 0$. Hence, the conditions of Lemma \ref{beta_convergence} are satisfied which implies that the following holds
\begin{equation}
\sum_{t=1}^T \frac{\beta_{t}}{t} < \infty.
\end{equation}
From the above equation it follows by the Kronecker's Lemma that \cite[Thm. 2.5.5]{Durrett}
\begin{equation}\label{beta_cesaro_mean}
\lim_{T \to \infty} \frac{1}{T}\sum_{t=1}^T \beta_{t}  = 0.
\end{equation}
The above convergence result implies that by Lemma 6 in \cite{Swenson_et_al_2014}, for any $\epsilon >0$, the number of centroid empirical frequencies away from the $\epsilon$ consensus NE is finite for any time $T$, that is,
\begin{equation}
\lim_{T\to \infty} \frac{\#\{1\leq t\leq T: \bar{f}_t^n \notin C_\epsilon(\mu)\}}{T} = 0. 
\end{equation}
The relation above implies that the distance between the empirical frequencies and the set of symmetric NE diminishes by Lemma \ref{convergence_consensus_equilibrium}, that is,
\begin{equation}
\lim_{t \to \infty} d( \bar{f}_t^n, C(\mu))=0.
\end{equation}
where $d(\cdot, \cdot)$ is the distance defined in \eqref{distance_definition}. The result in \eqref{eqn_average_empirical_convergence} follows from above. 
\end{myproof}

%
\section{Proof of Corollary \ref{consensus_nash_convergence}}\label{dfp_corollary_proof}

Denote the $n-1$ tuple of the average empirical distribution $\bar{f}_t$ by $\bar{f}_t^{n-1}$. By the Lipschitz continuity of the multilinear utility expectation we have that $\|u(a_i, \bar{f}_t^{n-1} ;\mu) - u(a_i, \nu^i_{-it} ;\mu_{it})\| \leq K \|(\bar{f}_t^{n-1}, \mu) - (\nu^i_{-it}, \mu_{it})\|$ for all $a_i$ where $\nu^i_{jt} = \hat{\bar{f}}^i_t$ for all $j \neq i$ and $K\geq0$ is the Lipschitz constant. By Lemma \ref{convergence_actions_beliefs} and Assumption \ref{state_learning_assumption}, we have $\|(\bar{f}_t^{n-1} ,\mu) - (\nu^i_{-it} ,\mu_{it})\| = O(\log t/t)$. Then using  \eqref{eqn_distinct_maximizer}, we have for all $a\in\ccalA\setminus a^*$
\begin{equation}
u(a^*,\nu^i_{-it} ;\mu_{it}) - u(a,\nu^i_{-it} ;\mu_{it}) \geq \epsilon - \delta_t
\end{equation}
for $\nu^i_{jt} = \hat{\bar{f}}^i_t$ for all $j \neq i$, $\delta_t \geq 0 $ and $\delta_t= O(\log t/t)$. Therefore, there exists a finite time $T>0$ such that $\epsilon - \delta_t>0$ for all $t>T$. This means that $a^*$ is the best response action of $i$ after time $T$. Then the empirical frequency of $i\in \ccalN$ $f_{it}$ converges to $\Psi(a^*)$ which implies \eqref{eqn_convergence_empirical_frequency}.

\section{Proof of Theorem \ref{histogram_sharing_convergence}}\label{dfp_histogram_sharing_appendix}

Before we prove the theorem, we first analyze the convergence rate of the histogram sharing presented in Section \ref{histogram_sharing_section} where we defined $\nu^i_{j t} = f_{jt}$ if $j \in \ccalN_i$ or $\nu^i_{jt}$ is given by \eqref{belief_update_non_neighbors} if $j \notin \ccalN_i$.

Denote the $l$th element of $\nu^i_{jt}$ by $\nu_{jt}^i(l)$. Define the matrix that captures population's estimate on $j$'s empirical distribution, $\hat{F}_{jt} := [\nu_{jt}^1,\dots,\nu_{jt}^n]^T \in \reals^{n \times m}$.  The $l$th column of $\hat{F}_{jt}$ represents the population's estimate on $j$'s $l$th local action denoted by $\hat{F}_{j t}(l) := [\nu_{jt}^1(l), \dots, \nu_{jt}^n(l)]^T \in \reals^{n \times 1}$. 

Observe that $j$'s estimate of the frequency of its own action $l$ is correct, that is, $\nu_{j t}^j(l) = f_{j t}(l)$. Define the vector $\bbx_{j l t} \in\reals^{n \times 1}$ where its $j$th element is equal to the empirical frequency of agent $j$ taking action $l\in \ccalA$, that is, $\bbx_{j l t}(j) = {f}_{j t}(l)$, and its other elements are zero. Further define the weighted adjacency matrix for belief update on the frequency of agent $j$'s $l$th action $W_{jl} \in \reals^{n \times n}$ with $W_{j l}(i,k) = w^i_{jk}$ for all $i$ and $k$. We remind that $w^i_{jk}$ is the weight that $i$ uses to mix agent $j\in \ccalN_i$'s belief on agent $k\notin \ccalN_i$'s empirical distribution in \eqref{belief_update_non_neighbors}. Also note that there are $m$ weight matrices $W_{j l}$ each corresponding to one action $l \in \ccalA$.

The matrix $W_{jl}$ is row stochastic, that is, the sum of row elements of $W_{jl}$ is equal to one for each row by  $\sum_{k\in\ccalN_i} w^i_{jk}= 1$ and we have that $W_{jl}(i,j) = 1$ for $j \in \ccalN_i \bigcup i$. The latter condition on $W_{jl}$ is by the fact that if $j\in \ccalN_i$, $j$ sends its action to its neighbor $i$ and hence $\nu^i_{jt} = f_{jt}$.  Given these definitions we can write a linear recursion for population's estimate of $j$'s empirical frequency of its $l$th action
\begin{equation} \label{estimate_frequency_agent_j_action_l}
\hat{F}_{j t+1}(l) = W_{j l} (\hat{F}_{j t}(l) + \bbx_{j l t+1} - \bbx_{j l t}).
\end{equation}

Note that if the above linear system converges to the true empirical frequency of ${f}_{j t}(l)$ in all of its elements then it implies that all agents learned its true value. 
%

Next, we analyze the linear update in \eqref{estimate_frequency_agent_j_action_l} and show the convergence of the belief of agent $i$ on the population's empirical distribution $\nu^i_{-it}$ to the true average empirical distribution of the rest of the population ${f}_{-it}$ at rate $O(\log t/t)$.
%
%
\begin{lemma} \label{convergence_histogram_sharing_fictitious_play}
Consider the distributed fictitious play in which the empirical distribution of agent $j$ $ f_{jt}$ evolves according to \eqref{empirical_distribution_recursion} and agent $i$ updates its estimate on the empirical play of the population $\nu^i_{-it}$ according \eqref{empirical_distribution_recursion} if $j\in \ccalN_i$ or using \eqref{belief_update_non_neighbors} if $j \notin \ccalN_i$. If the network satisfies Assumption \ref{connectivity_assumption} and the initial beliefs are the same for all agents, i.e., $\nu^i_{j1} = {f}_{j1}$ for all $i \in \ccalN$, then $\nu^i_{jt}$ converges in norm to $f_{jt}$ at the rate $O({\log t}/{t})$, that is, $\|\nu^i_{jt} - f_{jt}\| = O(\frac{\log t}{t})$ for all $j \in \ccalN$.
\end{lemma}

%
%

\begin{myproof}
We consider the difference between the population's estimate of the empirical frequency of $j$ taking action $l\in \ccalA$ and $j$'s true empirical distribution $f_{j t}(l) \bbone$ by subtracting $f_{j t+1}(l) \bbone$ from both sides of \eqref{estimate_frequency_agent_j_action_l} to get
\begin{equation}
\hat{F}_{j t+1}(l) - f_{j t+1}(l) \bbone = W_{j l} (\hat{F}_{j t}(l) + \bbx_{j l t+1} - \bbx_{j l t}) - f_{j t+1}(l) \bbone.
\end{equation}
Since $W_{j l}$ is row stochastic, we can move the last term on the right hand side inside the matrix multiplication,
\begin{equation}
\hat{F}_{j t+1}(l) - f_{j t+1}(l) \bbone = W_{j l} (\hat{F}_{j t}(l) + \bbx_{j l t+1} - \bbx_{j l t} -  f_{j t+1}(l) \bbone) .
\end{equation}
We can equivalently express $f_{j t+1}(l)  = f_{j t}(l) + \bbx_{j l t+1}(j) - \bbx_{j l t}(j)$ by the definition of the vector $\bbx_{j l t}$. Substituting this expression for the $f_{j t+1}(l)$ on the right hand side of the above equation we have 
\begin{align}
\hat{F}_{j t+1}(l) - &f_{j t+1}(l) \bbone = W_{j l} \Big(\hat{F}_{j t}(l) + \bbx_{j l t+1} - \bbx_{j l t} \nonumber \\
& - ( f_{j t}(l) + \bbx_{j l t+1}(j) - \bbx_{j l t}(j)) \bbone\Big).
\end{align}
Let $\bby_t := \hat{F}_{j t}(l) - f_{j t}(l) \bbone$, then
\begin{equation}\label{y_update}
\bby_{t+1} = W_{j l} (\bby_t +\bbx_{j l t+1} - \bbx_{j l t} - ( \bbx_{j l t+1}(j) - \bbx_{j l t}(j)) \bbone) .
\end{equation}

Let $\bbdelta_t := \bbx_{j l t+1} - \bbx_{j l t} - ( \bbx_{j l t+1}(j) - \bbx_{j l t}(j)) \bbone$. Next, we provide an upper bound for $\|\bbdelta_t\|$ by using the triangle inequality and observing the fact that recursion for fictitious play in \eqref{empirical_distribution_recursion} can change only the $j$th element of the vector $\bbx_{j l t}$ by $1/t$, that is, $\bbx_{j l t+1}(j) - \bbx_{j l t}(j) = \frac{1}{t} (\Psi(a_{j t})(l) - f_{j t}(l))$, as follows
\begin{align}
\|\bbdelta_t \| &= \|\bbx_{j l t+1} - \bbx_{j l t} - ( \bbx_{j l t+1}(j) - \bbx_{j l t}(j)) \bbone\|  \\
&\leq  \|\bbx_{j l t+1} - \bbx_{j l t} \| + \|\bbx_{j l t+1}(j) \bbone - \bbx_{j l t}(j) \bbone\| \\
& \leq \frac{1}{t} + \frac{n}{t} = \frac{n+1}{t} = O(\frac{1}{t}) \label{error_rate}.
\end{align}

Now consider the row stochastic matrix $W_{jl}$. Its largest eigenvalue is $\lambda_1 = 1$ and its right eigenvector is equal to column vector of ones $\bbone$ by the Perron-Frobenius theorem  \cite[Ch. 2.2]{Brouwer_Haemers_2011}. The left eigenvector associated with the eigenvalue $\lambda_1$ is given by $\bbe_j^T$. This is easy to see when we interpret $W_{jl}$ as representing a Markov chain where state $j$ is an absorbing state and there is a positive transition probability from any other state to state $j$. Note that once a state $i$ that is a neighbor of $j$ is reached, the transition to state $j$ is with probability 1 due to the update rule. Because the graph $\ccalG$ is strongly connected, for any $i\notin \ccalN_j$ there exists a path to a node $k\in \ccalN_j\bigcup j$. As a result the absorbing state $j$ is reached with positive probability which implies the stationary distribution of the Markov chain is given by $\bbe_{j}$, that is, with probability 1 the state is $j$. Moreover, $\lim_{t \to \infty} W_{jl}^t \to \bbone \bbe_j^T$. 

Now define the matrix $\overline{W}_{jl} = W_{jl} -\bbone \bbe_j^T$. By the fact that the limiting power sequence of the matrix is $\bbone \bbe_j^T$, $\lim_{t \to \infty} \overline{W}_{jl}^t \to \bbzero$.  
By its construction the sum of the row elements of $\overline{W}_{jl}$ is zero for any row, that is, $\overline{W}_{jl} \bbone = \bbzero_{n \times 1}$. Further note that the $j$th row of $\overline{W}_{jl}$ is all zeros as well as all the rows $k$ such that $j \in \ccalN_k$. 

By using the definition of $\bbdelta_t$, we can equivalently write \eqref{y_update} as
\begin{align}
\bby_{t+1} = W_{j l} (\bby_t + \bbdelta_t) =\sum_{s = 0}^t   W_{j l}^{s+1} \bbdelta_{t-s} +        W_{j l}^t \bby_1
\end{align}
for $t=1,2,\dots$.
The second equality follows by writing the first equality  for $\{\bby_s\}_{s=1,\dots,t}$ and iteratively substituting each term. Note that by the assumption $\nu^i_{j1} = {f}_{j1}$, $\bby_1 = \bbzero$. So when we consider the norm of $\bby_{t+1}$, $\|\bby_{t+1} \|$, we are left with
\begin{align}
\|\bby_{t+1}\| &= \|  \sum_{s = 0}^t   W_{j l}^{s+1} \bbdelta_{t-s}\|   \leq \sum_{s = 0}^t   \|W_{j l}^{s+1} \bbdelta_{t-s}\|
\end{align}
Now use the decomposition $W_{j l} = \overline{W}_{j l} + \bbone \bbe_j^T$ in the above line to get
\begin{equation}
\|\bby_{t+1}\| \leq \sum_{s = 0}^t   \|( \overline{W}_{j l} + \bbone \bbe_j^T)^{s+1} \bbdelta_{t-s}\|
\end{equation}
Since $\overline{W}_{j l} \bbone = \bbzero$, $\bbe_j^T \overline{W}_{j l} = \bbzero$ and  $  \bbone \bbe_j^T = (\bbone \bbe_j^T)^s$ for any $s = 1,2,\dots$, we have $W_{j l}^s =  \overline{W}_{j l}^s + \bbone \bbe_j^T$. Then we can upper bound $\|\bby_{t+1}\|$ by using the triangle inequality as follows
\begin{equation}
\|\bby_{t+1}\| \leq \sum_{s = 0}^t   \| \overline{W}_{j l}^{s+1} \bbdelta_{t-s} \| + \|(\bbone \bbe_j^T)^{s+1} \bbdelta_{t-s}\|
\end{equation}
Further note $\bbdelta_t(j) = 0$ for any $t=1,2,\dots$ by the definition of $\bbx_{jlt+1}$ and $\bbx_{jlt}$, and therefore $\bbe_j^T \bbdelta_t= 0$, which means the second term on the right hand side of the inequality is zero, that is,
\begin{equation}
\|\bby_{t+1}\| \leq \sum_{s = 0}^t   \| \overline{W}_{j l}^{s+1} \bbdelta_{t-s} \|. 
\end{equation}
Furthermore, the spectral radius of  $ \overline{W}_{jl}$ is strictly less than $1$, that is, $\bar\lambda_1 := \rho(\overline{W}_{jl}) <1$ because $\lim_{t \to \infty} \overline{W}_{jl}^t = \bbzero$ \cite[Thm. 1.10]{Varga_2009}. As a result, we have 
\begin{equation}
\|\bby_{t+1}\| \leq \sum_{s =0}^t   \| \overline{W}_{j l}^{s+1} \bbdelta_{t-s}\| \leq \sum_{s = 0}^t   \rho(\overline{W}_{j l})^{s+1} \| \bbdelta_{t-s}\|
\end{equation}
Note that by \eqref{error_rate}, we have $\|\bbdelta_{t-s}\| = n+1/t-s$. Define $\delta_{avg}(t) := \frac{1}{t}\sum_{s=1}^t \frac{n+1}{s}$. By Chebychev's sum inequality \cite{Hardy_et_al} (p. 43-44), we have 
\begin{align}
\|\bby_{t+1}\| &\leq  \delta_{avg}(t) \sum_{s = 0}^t   \bar\lambda_1^{s+1} \\
& =  \delta_{avg}(t) (\bar\lambda_1 \frac{1-\bar\lambda_1^{t+1}}{1-\bar\lambda_1})\leq \frac{\delta_{avg}(t)}{1-\bar\lambda_1}
\end{align}
Noting that $\delta_{avg}(t) := \frac{1}{t}\sum_{s=1}^t \frac{n+1}{s} = O(\frac{\log t}{t})$, we have $\|\bby_{t+1}\| = \| \hat{F}_{j t}(l) - f_{j t}(l) \bbone\| =  O(\frac{\log t}{t})$ for any $l \in \ccalA$. Consequently, $\|\nu_{jt}^i - f_{jt}\| = O(\frac{\log t}{t})$. 
%
\end{myproof}

Similar to Lemma \ref{convergence_actions_beliefs} the above result is true irrespective of the game that the agents are playing. The result leverages on the fact that the change in the empirical distribution of agent $j$ is at most $1/t$ by the recursion in \eqref{empirical_distribution_recursion} and the belief updates of $i$ on $j$'s empirical frequency evolves faster than the change in agent $j$'s empirical distribution. We continue with the proof of the Theorem.

%
%
\begin{myproof}[ of Theorem \ref{histogram_sharing_convergence}]
Proof follows the same proof outline in Theorem \ref{action_sharing_convergence}. Start by exploiting the multi-linearity of the expected utility when all individuals play with respect to their empirical distributions \cite{Monderer_Shapley}, that is, 
\begin{align}
u({f}_{t+1}; \mu&) =  u({f}_{t}; \mu)  \nonumber \\
&+ \frac{1}{t} \sum_{i =1}^n u(\Psi(a_{it}), {f}_{-i t}; \mu) - u({f}_{it}, {f}_{-it};\mu)+  \frac{\delta}{t^2}.
\end{align}
for some $\delta >0$ which we collect higher order terms. 
We move the first term of the RHS to the left and add $|\delta|/t^2$ to the left hand side and get rid of the last term on the right hand side,
\begin{align}
u({f}_{t+1};&\, \mu) - u({f}_{t}; \mu) + \frac{|\delta|}{t^2} \geq \nonumber \\
&\frac{1}{t} \sum_{i =1}^n u(\Psi(a_{it}), {f}_{-i t}; \mu) - u({f}_{it}, {f}_{-it}; \mu).  
\end{align}
Now define $L_{it} : = v(\nu^i_{-it}; \mu_{i t})-u(\Psi(a_{it}), f_{-it}; \mu)$. Add $\sum_{i=1}^n L_{it} /t$ to both sides of the above equation to get
\begin{align}
u({f}_{t+1};\, &\mu) - u({f}_{t}; \mu) + \frac{|\delta|}{t^2} + \frac{1}{t}\sum_{i=1}^n L_{it}  \nonumber \\
&\geq \frac{1}{t} \sum_{i =1}^n v(\nu^i_{-it}; \mu_{i t}) - u({f}_{it}, {f}_{-it};\mu) . 
\end{align}
Now we sum up the terms above from time $t=1$ to $T$,
\begin{align} \label{bound_best_response_over_time}
u({f}_{T+1}; &\,\mu) - u({f}_{1}; \mu)  + \sum_{t=1}^{T+1} \frac{|\delta|}{t^2} +  \sum_{t=1}^{T+1} \frac{1}{t}\sum_{i=1}^n L_{it}\nonumber\\
& \geq  \sum_{t=1}^{T+1} \frac{1}{t} \sum_{i =1}^n v(\nu^i_{-it}; \mu_{i t}) - u({f}_{it}, {f}_{-it};\mu). 
\end{align}
Consider the left hand side of the above equation. The utility and therefore the expected utility is bounded. The third term is summable. By Lemma \ref{convergence_histogram_sharing_fictitious_play} and Assumption \ref{state_learning_assumption}, the conditions of Lemma \ref{L_convergence} are satisfied. Lemma \ref{L_convergence} yields that the last term on the left hand side of \eqref{bound_best_response_over_time} is summable. Hence, the left hand side of \eqref{bound_best_response_over_time} is bounded. 

Define $\alpha_{t} :=\sum_{i =1}^n v(\nu^i_{-it}; \mu_{i t}) - u({f}_{it}, {f}_{-it}; \mu)$. Using the definition of $\alpha_{t}$ and the boundedness of the left hand side of the above equation, it follows from \eqref{bound_best_response_over_time} that there exists some bounded parameter $0<\bar{B}< \infty$ such that 
\begin{align}
\bar{B} >  \sum_{t=1}^{\infty} \frac{\alpha_t}{t}.
\end{align}
Define $\beta_{t}:= \sum_{i=1}^n v(f_{-it}; \mu) - u(f_{it}, f_{-it}; \mu)$ and consider the difference between $\alpha_{t+1}$ and $\beta_{t+1}$ 
\begin{align}
\|\alpha_{t} - \beta_{t}\| = \| \sum_{i = 1}^n v(\nu^i_{-it}; \mu_{i t}) - v(f_{-it}; \mu)\|
\end{align}
Lemma \ref{L_convergence} implies that the above equality is equal to $\|\alpha_{t} - \beta_{t}\| = O(\log t/t)$. By noting that $\beta_{t} \geq 0$, the conditions of Lemma \ref{beta_convergence} are satisfied which implies that 
\begin{align}
\sum_{t=1}^{T} \frac{\beta_{t}}{t} < \infty
\end{align}
for any $T >0$. As a result the time average of the above sum converges to zero by Kronecker's Lemma \cite[Thm. 2.5.5]{Durrett}, that is, 
\begin{align}
\lim_{T\to \infty} \frac{1}{T}\sum_{t=1}^{T} \frac{\beta_{t}}{t}  = 0.
\end{align}
We remark that $\beta_t$ captures the difference in expected payoffs when agent $i$ best responds to others' empirical distribution $f_{-it}$ given the common asymptotic belief $\mu$, and when agent $i$ follows its own empirical distribution $f_{it}$ with common beliefs on the state $\mu$. 
The convergence in \eqref{histogram_convergence} follows from the above equation by Lemma \ref{convergence_equilibrium}.
\end{myproof}
%

\end{appendices}

\bibliographystyle{IEEEtran}
\bibliography{bibliography}

\begin{thebibliography}{10}
\providecommand{\url}[1]{#1}
\csname url@samestyle\endcsname
\providecommand{\newblock}{\relax}
\providecommand{\bibinfo}[2]{#2}
\providecommand{\BIBentrySTDinterwordspacing}{\spaceskip=0pt\relax}
\providecommand{\BIBentryALTinterwordstretchfactor}{4}
\providecommand{\BIBentryALTinterwordspacing}{\spaceskip=\fontdimen2\font plus
\BIBentryALTinterwordstretchfactor\fontdimen3\font minus
  \fontdimen4\font\relax}
\providecommand{\BIBforeignlanguage}[2]{{%
\expandafter\ifx\csname l@#1\endcsname\relax
\typeout{** WARNING: IEEEtran.bst: No hyphenation pattern has been}%
\typeout{** loaded for the language `#1'. Using the pattern for}%
\typeout{** the default language instead.}%
\else
\language=\csname l@#1\endcsname
\fi
#2}}
\providecommand{\BIBdecl}{\relax}
\BIBdecl

\bibitem{Shoham_Leyton_2008}
Y.~Shoham and K.~Leyton-Brown, \emph{Multiagent systems: Algorithmic,
  game-theoretic, and logical foundations}.\hskip 1em plus 0.5em minus
  0.4em\relax Cambridge University Press, 2008, vol.~1.

\bibitem{EksinRibeiro_2012}
C.~Eksin and A.~Ribeiro, ``Distributed network optimization with heuristic
  rational agents,'' \emph{IEEE Trans. Signal Process.}, vol.~60, no.~10, pp.
  5396--5411, October 2012.

\bibitem{NedicOzdaglar}
A.~Nedic and A.~Ozdaglar, ``Distributed subgradient methods for multiagent
  optimization,'' \emph{IEEE Trans. Autom. Control}, vol.~54, no.~1, 2009.

\bibitem{Tsitsiklisetal}
J.~Tsitsiklis, D.~Bertsekas, and M.~Athans, ``Distributed asynchronous
  deterministic and stochastic gradient optimization algorithms,'' \emph{IEEE
  Trans. Autom. Control}, vol.~31, no.~9, pp. 803--812, 1986.

\bibitem{chenSayed}
J.~Chen and A.~Sayed, ``Diffusion adaptation strategies for distributed
  optimization and learning over networks,'' \emph{Signal Processing, IEEE
  Transactions on}, vol.~60, no.~8, pp. 4289--4305, 2012.

\bibitem{Monderer_Shapley_1996a}
D.~Monderer and L.~Shapley, ``Fictitious play property for games with identical
  interests,'' \emph{Journal of economic theory}, vol.~68, no.~1, pp. 258--265,
  1996.

\bibitem{Swenson_et_al_2014}
B.~Swenson, S.~Kar, and J.~Xavier, ``Empirical centroid fictitious play: An
  approach for distributed learning in multi-agent games,'' \emph{IEEE Trans.
  Signal Process.}, vol.~PP, no.~99, p.~1, 2015.

\bibitem{Marden_et_al_2009}
J.~Marden, G.~Arslan, and J.~Shamma, ``Joint strategy fictitious play with
  inertia for potential games,'' \emph{IEEE Trans. Automatic Control}, vol.~54,
  no.~2, pp. 208--220, 2009.

\bibitem{Shamma_Arslan_2005}
J.~Shamma and G.~Arslan, ``Dynamic fictitious play, dynamic gradient play, and
  distributed convergence to nash equilibria,'' \emph{IEEE Trans. Automatic
  Control}, vol.~50, no.~3, pp. 312--327, 2005.

\bibitem{Hart_2005}
S.~Hart, ``Adaptive heuristics,'' \emph{Econometrica}, vol.~73, no.~5, pp.
  1401--1430, 2005.

\bibitem{Young_2004}
H.~Young, \emph{Strategic learning and its limits}.\hskip 1em plus 0.5em minus
  0.4em\relax Oxford University Press, 2004.

\bibitem{Harsanyi_1968}
J.~Harsanyi, ``Games with incomplete information played by bayesian players -
  part ii. bayesian equilibrium points.'' \emph{Management Science}, vol.~14,
  no.~5, pp. 320--334, 1968.

\bibitem{EksinEtal13_b}
C.~Eksin, P.~Molavi, A.~Ribeiro, and A.~Jadbabaie, ``Learning in networks with
  incomplete information:asymptotic analysis and tractable implementation of
  rational behavior,'' \emph{IEEE Signal Process. Mag.}, vol.~30, no.~3, pp.
  30--42, May 2013.

\bibitem{Eksin_et_al_2013}
------, ``Bayesian quadratic network game filters,'' \emph{IEEE Trans. Signal
  Process.}, vol.~62, no.~9, pp. 2250 -- 2264, May 2014.

\bibitem{Dekel_et_al_2004}
E.~Dekel, D.~Fudenberg, and D.~Levine, ``Learning to play bayesian games,''
  \emph{Games and Economic Behavior}, vol.~46, no.~2, pp. 282--303, 2004.

\bibitem{Fudenberg_Levine_1998}
D.~Fudenberg and D.~Levine, \emph{The Theory of Learning in Games},
  1st~ed.\hskip 1em plus 0.5em minus 0.4em\relax Cambridge, MA: MIT Press,
  1998.

\bibitem{Brown_1951}
G.~W. Brown, ``Iterative solution of games by fictitious play,'' \emph{Activity
  analysis of production and allocation}, vol.~13, no.~1, pp. 374--376, 1951.

\bibitem{Fudenberg_Kreps_1993}
D.~Fudenberg and D.~Kreps, ``Learning mixed equilibria,'' \emph{Games and
  Economic Behavior}, vol.~5, no.~3, pp. 320--367, 1993.

\bibitem{Fudenberg_Takahashi}
D.~Fudenberg and S.~Takahashi, ``Heterogeneous beliefs and local information in
  stochastic fictitious play,'' \emph{Games and Economic Behavior}, vol.~71,
  no.~1, pp. 100--120, 2011.

\bibitem{Arslan_et_al_2007}
G.~Arslan, J.~Marden, and J.~Shamma, ``Autonomous vehicle-target assignment: A
  game-theoretical formulation,'' \emph{Journal of Dynamic Systems,
  Measurement, and Control}, vol. 129, no.~5, pp. 584--596, 2007.

\bibitem{Marden_Shamma_2012}
J.~Marden and J.~Shamma, ``Revisiting log-linear learning: Asynchrony,
  completeness and a payoff-based implementation,'' \emph{Games and Economic
  Behavior}, vol.~75, no.~2, pp. 788--808, 2012.

\bibitem{Jadbabaie_et_al_2003}
A.~Jadbabaie, J.~Lin, and A.~Morse, ``Coordination of groups of mobile
  autonomous agents using nearest neighbor rules,'' \emph{IEEE Trans. Autom.
  Control}, vol.~48, no.~6, pp. 988--1001, 2003.

\bibitem{Kashyap_et_al_2007}
A.~Kashyap, T.~Basar, and R.~Srikant, ``Quantized consensus,''
  \emph{Automatica}, vol.~43, no.~7, pp. 1192--1203, 2007.

\bibitem{Ribeiro_ConsensusI}
I.~Schizas, A.~Ribeiro, and G.~Giannakis, ``Consensus in ad hoc wsns with noisy
  links - part i: distributed estimation of deterministic signals,'' \emph{IEEE
  Trans. Signal Process.}, vol.~56, no.~1, pp. 1650--1666, January 2008.

\bibitem{StankovicStankovic}
S.~Stankovic, M.~Stankovic, and D.~Stipanovic, ``{Decentralized parameter
  estimation by consensus based stochastic approximation},'' in \emph{{Proc. of
  the 46th IEEE Conference on Decision and Control (CDC)}}, New Orleans, LA,
  USA, Dec. 2007, pp. 1535--1540.

\bibitem{olfati2007distributed}
R.~Olfati-Saber, ``Distributed kalman filtering for sensor networks,'' in
  \emph{46th IEEE Conference on Decision and Control, 2007}.\hskip 1em plus
  0.5em minus 0.4em\relax IEEE, 2007, pp. 5492--5498.

\bibitem{Olfati_et_al_2006}
R.~Olfati-Saber, E.~Franco, E.~Frazzoli, and J.~S. Shamma, ``Belief consensus
  and distributed hypothesis testing in sensor networks,'' in \emph{Networked
  Embedded Sensing and Control}.\hskip 1em plus 0.5em minus 0.4em\relax
  Springer Berlin Heidelberg, 2006, pp. 169--182.

\bibitem{SoummyaKar_a}
S.~Kar, J.~M. Moura, and K.~Ramanan, ``Distributed parameter estimation in
  sensor networks: Nonlinear observation models and imperfect communication,''
  \emph{IEEE Tran. Information Theory}, vol.~58, no.~6, pp. 3575--3605, 2012.

\bibitem{Shahrampour_et_al_2014}
S.~Shahrampour, A.~Rakhlin, and A.~Jadbabaie, ``Distributed detection:
  Finite-time analysis and impact of network topology,'' 2014.

\bibitem{Jadbabaie_et_al_2013}
A.~Jadbabaie, P.~Molavi, and A.~Tahbaz-Salehi, ``Information heterogeneity and
  the speed of learning in social networks,'' \emph{Columbia Business School
  Research Paper}, pp. 13--28, 2013.

\bibitem{Vives_1997}
X.~Vives, ``Learning from others: a welfare analysis.'' \emph{Games Econ.
  Behav.}, vol.~20, no.~2, pp. 177--200, 1997.

\bibitem{Gale_Kariv}
D.~Gale and S.~Kariv, ``Bayesian learning in social networks,'' \emph{Games
  Econ. Behav.}, vol.~45, no.~2, pp. 329--346, 2003.

\bibitem{Djuric_2012}
P.~Djuric and Y.~Wang, ``Distributed bayesian learning in multiagent systems,''
  \emph{IEEE Signal Process. Mag.}, vol.~29, pp. 65--76, March, 2012.

\bibitem{Mueller_2013}
M.~Mueller-Frank, ``A general framework for rational learning in social
  networks,'' \emph{The Theoretical Economics}, vol.~8, pp. 1--40, 2013.

\bibitem{Nedic_et_al_2009}
A.~Nedic, A.~Olshevsky, A.~Ozdaglar, and J.~Tsitsiklis, ``On distributed
  averaging algorithms and quantization effects,'' \emph{IEEE Trans. Autom.
  Control}, vol.~54, no.~11, 2009.

\bibitem{Monderer_Shapley}
D.~Monderer and L.~Shapley, ``Fictitious play property for games with identical
  interests,'' \emph{Journal of economic theory}, vol.~68, no.~1, pp. 258--265,
  1996.

\bibitem{Durrett}
R.~Durrett, \emph{Probability: Theory and Examples}, 3rd~ed.\hskip 1em plus
  0.5em minus 0.4em\relax Cambridge Series in Statistical and Probabilistic
  Mathematics, 2005.

\bibitem{Brouwer_Haemers_2011}
A.~E. Brouwer and W.~H. Haemers, \emph{Spectra of graphs}.\hskip 1em plus 0.5em
  minus 0.4em\relax Springer, 2011.

\bibitem{Varga_2009}
R.~S. Varga, \emph{Matrix iterative analysis}.\hskip 1em plus 0.5em minus
  0.4em\relax Springer Science \& Business, 2009, vol.~27.

\bibitem{Hardy_et_al}
G.~H. Hardy, J.~E. Littlewood, and G.~Polya, \emph{Inequalities, Cambridge
  Mathematical Library}.\hskip 1em plus 0.5em minus 0.4em\relax Cambridge
  University Press, 1988.

\end{thebibliography}

\end{document}